\documentclass[11pt,cite,a4paper]{article}

\pdfoutput=1

\usepackage{amsmath}
\usepackage{amssymb}
\usepackage{a4wide}
\usepackage{graphicx}

\numberwithin{equation}{section}

\begin{document}

\begin{titlepage}

\rightline{MIT-CTP 4188}
\vspace{2.5truecm}

\centerline{\Large \bf  Holographic entanglement entropy:}
\vspace{.2cm}
\centerline{\Large \bf  near horizon geometry and disconnected regions}

\vspace{1.6cm}

\centerline{Erik Tonni}

\vspace{1.2cm}

\centerline{{\it  Center for Theoretical Physics,}}
\vspace{.1cm}
\centerline{{\it  Massachusetts Institute of Technology,}}
\vspace{.1cm}
\centerline{{\it Cambridge, MA 02139, USA}}

\vspace{.5cm}
\centerline{{\tt tonni@mit.edu}}

\vspace{2.5truecm}

\centerline{\bf Abstract}
\vspace{.5truecm}
We study the finite term of the holographic entanglement entropy for the charged black hole in $AdS_{d+2}$ and other examples of black holes when the spatial region in the boundary theory is given by one or two parallel strips. For one large strip it scales like the width of the strip.  The divergent term of its   expansion as the turning point of the minimal surface approaches the horizon is determined by the near horizon geometry.
Examples involving a Lifshitz scaling are also considered.
For two equal strips in the boundary we study the transition of the mutual information given by the holographic prescription. 
In the case of the charged black hole, when the width of the strips becomes large this transition provides a characteristic finite distance depending on the temperature.

\vspace{2truecm}

\end{titlepage}

\section*{Introduction}

Entanglement entropy is an important quantity which has been studied in many models of condensed matter systems, quantum information and quantum gravity. It measures the quantum correlations in a bipartite decomposition of a quantum system.\\
Let us consider a system whose total Hilbert space can be written as a direct product $H=H_A \otimes H_B$. 
Denoting by $\rho$ the density matrix characterizing the state of the system, the reduced density matrix associated to $A$ is obtained by tracing $\rho$ over the degrees of freedom of $B$, i.e. $\rho_A = \textrm{Tr}_B \rho$. Then, the entanglement entropy is defined as the corresponding Von Neumann entropy, namely $S_A = -  \,\textrm{Tr}_A(\rho_A \log \rho_A)$. When the system is in a pure state, we have $\rho = | \Psi \rangle \langle \Psi |$ and $S_A = S_B$.
A very interesting situation occurs when $A$ and $B$ correspond to a spatial bipartition of the system. In this case the entanglement entropy is called geometric entropy. Here we consider this quantity and we will always refer to it as the  entanglement entropy. Its interesting feature is that is satisfies the so called area law: the leading term in the expansion for small UV cutoff $a$ is proportional to the area of the boundary separating $A$ and $B$. In $d$ spatial dimensions we have $S_A \propto \textrm{Area}(\partial A)/a^{d-1}+ \dots$, where the dots represent higher order terms in $a$ \cite{Srednicki:1993im}. This area law is violated in two dimensional conformal field theories, where a logarithmic behavior has been found for one interval. In particular $S_A =(c/3)\log(\ell/a)$ where $\ell$ is the length of the interval and $c$ is the central charge of the theory. The method employed to get the analytic result for $S_A$ is the replica trick, which means first to compute $\textrm{Tr} \rho_A^n$ for integer $n$ and then to perform an analytic continuation to real values of $n$ in order to take $S_A= -\,\partial_n \textrm{Tr} \rho_A^n \big|_{n=1}$  \cite{Callan:1994py,Holzhey:1994we,Calabrese:2004eu} (see \cite{Calabrese:2009qy} for a recent review).\\
For quantum field theories with a holographic dual, the problem of computing the entanglement entropy through a bulk description has been addressed in \cite{Ryu:2006bv,Ryu:2006ef}. The holographic prescription to obtain $S_A$ associated to a region $A$ in the $d+1$ dimensional boundary theory is the following.
On a fixed time slice (see \cite{Hubeny:2007xt} for a generalization to time dependent backgrounds), among all the $d$ dimensional surfaces extended in the bulk whose boundary coincides with the boundary of $A$, we have to consider the one having minimal area. Denoting this minimal surface by $\gamma_A$, the holographic entanglement entropy is given by $S_A=\textrm{Area}(\gamma_A)/(4 G_N^{(d+2)})$, where $G_N^{(d+2)}$ is the Newton constant of the $d+2$ dimensional theory in the bulk. Besides recovering the area law, this prescription passed many tests (e.g. the strong subadditivity inequalities) and it has been deeply studied (see the recent review \cite{Nishioka:2009un} and the references therein); thus it is considered a key tool to understand the essential features of the entanglement entropy for quantum field theories with a holographic dual.\\
The entanglement entropy is not an extensive quantity, as can be easily understood e.g. by the fact that $S_A =S_B$ for a pure state (this equality is violated at finite temperature). In the holographic computation of the entanglement entropy extensivity is recovered if one considers the finite term of the minimal area (sometimes called renormalized entanglement entropy), i.e. the one obtained by subtracting the UV divergent term giving the area law, in a black hole background \cite{Ryu:2006ef, Klebanov:2007ws, Barbon:2008sr, Barbon:2008ut}. This behavior is due to the fact that, as the size of the region $A$ in the boundary tends to infinity, a large part of the minimal surface gets very close to the horizon and this part goes like the volume of $A$ in the large size limit. Thus the near horizon geometry is responsible of the leading divergence of the finite term of the minimal area as the turning point of the minimal surface approaches the horizon.\\
A second important aspect of the entanglement entropy we are interested in concerns the case of a spatial region $A$ in the boundary made by two disjoint regions, i.e. $A= A_1 \cup A_2$ with $ A_1 \cap A_2 = \emptyset$. In this case the natural quantity to consider is the mutual information $M_A \equiv S_{A_1}+S_{A_2}-S_{A_1 \cup A_2}$ because it is UV finite. For some spin chains and two dimensional conformal field theories interesting results have been obtained \cite{Caraglio:2008pk, Furukawa:2008uk, Calabrese:2009ez, Alba:2009ek, Fagotti:2010yr, Casini:2005rm, Casini:2008wt, Casini:2009vk}. The models considered in these papers have small central charges (order of the unity).\\
In the context of the holographic correspondence, the case of disjoint regions has been addressed in \cite{Ryu:2006bv,Ryu:2006ef, Hirata:2006jx, Headrick:2007km, Hubeny:2007re, Headrick:2010zt}. An interesting feature of the holographic entanglement entropy is the transition of the mutual information from zero value to a positive value (the mutual information cannot be negative, as a consequence a strong subadditivity inequality) \cite{Hirata:2006jx, Headrick:2010zt}. This transition should be a large $c$ effect, which is the regime where the holographic prescription works, since there no signal of it e.g. for the compactified boson \cite{Furukawa:2008uk, Calabrese:2009ez}, which has $c=1$.\\
In this paper we consider the holographic entanglement entropy for one or two strips in the boundary theory in presence of various types of black holes with non compact horizon in the bulk. 
For one strip, we focus on the divergence of the finite term when the strip becomes large and therefore the turning point of the minimal surface approaches the horizon. The degree of this divergence depends on the near horizon geometry, but the finite term scales like the width (and thus like the volume) of the strip for all the black holes we consider. This scaling is broken for the Lifshitz type backgrounds whose dynamical exponent occurs in the spatial part of the metric.\\
For two parallel strips of equal width we mainly consider the transition of the mutual information in terms of the geometrical parameters, namely the width of the strips and the distance between them.  For the charged black hole in four dimensions with fixed charge, we find that the transition of the mutual information leads to a characteristic finite distance between the strips as they become large. This distance depends on the temperature and it could be interpreted as a signal of the occurrence of a finite correlation length in the boundary theory.\\
The paper is organized as follows. In the section \ref{section strip BH f(z)} we review the holographic prescription for the entanglement entropy, specializing the analysis to ansatz that contain the black hole metrics we consider in the rest of the paper. In the section \ref{section near horizon} we study the finite term of the holographic entanglement entropy for the charged black hole, the warped black hole of \cite{Herzog:2009gd} and the perturbed Lifshitz background considered in \cite{Gubser:2009cg} as a solution of the Abelian Higgs model \cite{Gubser:2008px}. 
In the section \ref{section lifshitz bh} we study the Lifshitz black hole of \cite{Balasubramanian:2009rx} computing the analytic expression of the holographic entanglement entropy to all orders in the UV cutoff. This allows us to extract the finite term and to test the method employed for the other black holes. In the section \ref{section 2 strips} we consider two equal and parallel strips in the boundary and study the transition of the mutual information for $AdS_{d+2}$ and for the charged black hole.

\section{Holographic entanglement entropy for black holes}
\label{section strip BH f(z)}

\begin{figure}[h]
\begin{center}
\includegraphics[width=10cm]{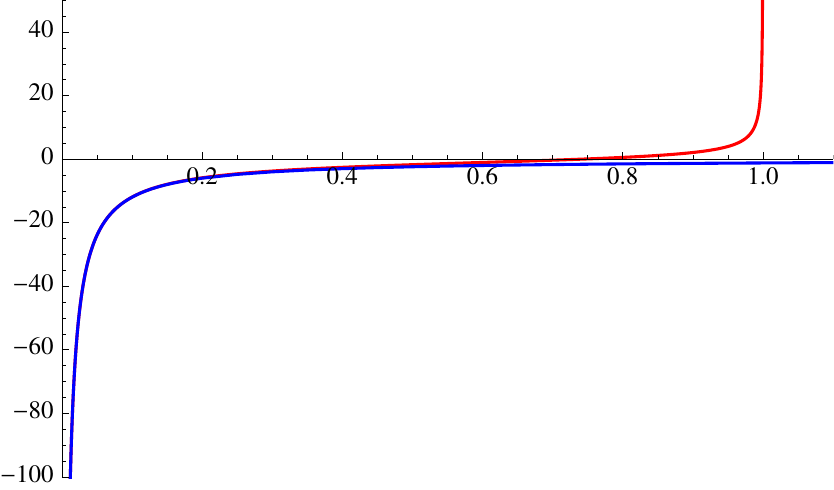}
\end{center}
\caption{Charged black hole, extremal case and $z_0=1$. Plot of the finite term $\mathcal{A}_2(z_{\textrm{max}},0)$ as function of $z_{\textrm{max}}$ (red curve). Close to the boundary (i.e. when $z_{\textrm{max}} \rightarrow 0$) it coincides with the curve corresponding to $AdS_4$  (blue curve), which can be read from (\ref{A expansion empty}).
\label{plotA0}}
\end{figure}

In this section we review the holographic prescription to compute the entanglement entropy \cite{Ryu:2006bv,Ryu:2006ef}, defining the integrals we need in order to study the black holes that we will consider in the remaining sections. In the appendix \ref{app AdS} we review the results for $AdS_{d+2}$, that will be also employed in the section \ref{section 2 strips}.\\
Given a quantum field theory living on the boundary $\mathbb{R}_t \times \mathbb{R}^d$ of an asymptotically $AdS_{d+2}$ space, we take a $d$ dimensional region $A$ strictly included in the constant time slice of the boundary. 
Let us take a $d$-dimensional surface $\gamma$ embedded in the constant time slice of $AdS_{d+2}$ defined by $z = z(\vec{x})$, being $z$ is the holographic coordinate and $\vec{x} \in \mathbb{R}^d$ a vector of the constant time section of the boundary. The area of $\gamma$ reads
\begin{equation}
\textrm{Area}(\gamma) = \int dx_1 \dots dx_d \,\sqrt{\textrm{det}(h_{ij})}
\end{equation}
where $h_{ij}$ is the induced metric on the surface $ds^2_{\textrm{ind}} =h_{ij} dx^i dx^j$.
Among all these surfaces, we restrict our attention to the ones whose boundary coincides with the boundary of the region $A$. Within this smaller subset of surfaces, we denote by $\gamma_A$ the one having minimal area.\\
The proposal of Ryu and Takayanagi \cite{Ryu:2006bv, Ryu:2006ef} is that we can holographically  compute the entanglement entropy in the boundary theory through a computation in the bulk. In particular
\begin{equation}
\label{RT prescription}
S_A = \frac{\textrm{Area}(\gamma_A)}{4G_N^{(d+2)}}
\end{equation}
where $G_N^{(d+2)}$ is the Newton constant in $d+2$ spacetime dimensions.\\
Depending on the shape of $\partial A$, one decides if it is more convenient to work either in cartesian ($d\vec{x}^2=\sum_{i=1}^d (dx^i)^2$) or polar coordinates ($d \vec{x}^2=d\rho^2+\rho^2 d\Omega^2_{d-1}$, being $d\Omega^2_{d-1}$ the metric of the $d-1$ dimensional unit sphere) of $\mathbb{R}^d$. Since we will mostly consider $A$ to be a finite strip or a disjoint union of two of them, we will adopt the cartesian coordinates for $\mathbb{R}^d$ (for an example where the polar coordinates system is employed, see the appendix \ref{app circle}, which contains a discussion on the circular case in the black hole background considered below).\\
For many known black holes which are asymptotically $AdS_{d+2}$, the metric on the fixed time slice is given by
\begin{equation}
\label{ds20 ansatz}
ds^2_0 \equiv
ds^2 \big|_{t = \textrm{const}} 
= R^2\left(
\frac{d\vec{x}^2}{z^2}+\frac{dz^2}{z^2f(z)}\right)
\end{equation}
where $R$ is the radius of $AdS_{d+2}$ realized closed to the boundary.
In this system of coordinates the boundary is the $z=0$ slice and the horizon is characterized by the smallest zero of the emblacking function $f(z)$. \\
Let us consider the region $A$ in the boundary given by a strip with length $L$ along one direction, that we call $x$, and $L_\perp$ along the other orthogonal ones.
Choosing the origin in the center of this strip, the symmetry of the problem allows us to restrict to surfaces described by the even function $z=z(x)$. Then, the area functional that we have to minimize to compute the holographic entanglement entropy reads
\begin{equation}
\label{strip area0}
\textrm{Area}(\gamma_A)
=
2R^d L_\perp^{d-1} \int_{0}^{\frac{L}{2}} 
dx \, \frac{1}{z^d}\,\sqrt{1+\frac{(z')^2}{f(z)}}\:.
\end{equation}
Considering as a Lagrangian density $\mathcal{L}_{\textrm{strip}}[z(x)]$ the integrand in (\ref{strip area0}), one notices that it does not depend explicitly on $x$. This is the main simplification that makes the case of a rectangular region $A$ easier to solve than the case of a circular region.
Indeed, the independence of $\mathcal{L}_{\textrm{strip}}$ on $x$ leads to the conserved quantity $\mathcal{H}_{\textrm{strip}}\equiv p_z z'-\mathcal{L}_{\textrm{strip}}$, where $p_z \equiv \partial \mathcal{L}_{\textrm{strip}}/\partial z'$. In particular, one gets $\mathcal{H}_{\textrm{strip}} = z^{-d}[1+(z')^2/f(z)]^{-1/2}$. By introducing $z_{\textrm{max}}^{2d}\equiv 1/\mathcal{H}_{\textrm{strip}}^2$, the constancy of $\mathcal{H}_{\textrm{strip}}$ reads
\begin{equation}
\label{zprime^2 strip}
z'\,=\,- \sqrt{f(z)}\,\frac{\sqrt{z_{\textrm{max}}^{2d}-z^{2d}}}{z^{d}}
\end{equation}
where we have used that $z' <0$.
This equation tells us that $z_{\textrm{max}}$ is the turning point, namely $z'=0$ when $z=z_{\textrm{max}}$. Notice that, from (\ref{zprime^2 strip}), also at the horizon $z_0$ we could have $z'=0$ because $f(z_0)=0$, but we never reach it because $z_0 > z_{\textrm{max}}\geqslant z(x) \geqslant 0$.
The equation (\ref{zprime^2 strip}) provides the profile of the minimal surface we are looking for and, by construction, it satisfies $z(L/2)=0$ and $z(0)=z_{\textrm{max}}$. As a check of (\ref{zprime^2 strip}), one can write the equation of motion coming from $\mathcal{L}_{\textrm{strip}}$
\begin{equation}
\frac{z'' z}{f(z)}
+d \left[1+\frac{(z')^2}{f(z)}\right]
- \frac{(z')^2 z}{2 f(z)^2}\,f'(z)
\,=\,0
\end{equation}
and verify that the same equation can be found by deriving the conservation law $\mathcal{H}_{\textrm{strip}}= \textrm{const}$ w.r.t. $x$.
Then, separating the variables in (\ref{zprime^2 strip}), we find that the inverse function $x(z)$ is 
\begin{equation}
\label{x(z)}
x(z)\,=\,\int_0^x d\tilde{x}\,=\,- \int_{z_{\textrm{max}}}^{z}
\frac{w^d}{\sqrt{f(w)}\,\sqrt{z_{\textrm{max}}^{2d}-w^{2d}}}\,dw\:.
\end{equation}
Imposing in (\ref{x(z)}) the relation $x(0)=L/2$, one gets that 
\begin{equation}
\label{L(zmax)}
\frac{L}{2}\,=\,
\int_0^{z_{\textrm{max}}}
\frac{w^d}{\sqrt{f(w)}\,\sqrt{z_{\textrm{max}}^{2d}-w^{2d}}}\,dw
\end{equation}
which provides $L=L(z_{\textrm{max}})$ and the correspondence between $z_{\textrm{max}}$ and $L$.\\
As for the area of the minimal surface defined by (\ref{zprime^2 strip}), we can employ its definition to change integration variable in (\ref{strip area0}), which therefore becomes 
\begin{equation}
\label{strip area1}
\textrm{Area}(\gamma_A)
= 2R^dL_\perp^{d-1} \int_{0}^{z_{\textrm{max}}} 
\frac{z_{\textrm{max}}^d}{z^d \sqrt{f(z)}\sqrt{z_{\textrm{max}}^{2d}-z^{2d}}}\,dz\:.
\end{equation}
It is important to remark that, since $f(z) \rightarrow 1$ as $z \rightarrow 0$, the integral in (\ref{strip area1}) diverges at $z=0$.
This leads us to put a UV cutoff  $z \geqslant a$ in the integration domain of (\ref{strip area1}). Thus, the integral we have to compute reads
\begin{equation}
\label{strip area}
\textrm{Area}(\gamma_A)
\,=\,2R^dL_\perp^{d-1} \int_a^{z_{\textrm{max}}} 
\frac{z_{\textrm{max}}^d}{w^d \sqrt{f(w)}\sqrt{z_{\textrm{max}}^{2d}-w^{2d}}}\,dw
\,\equiv\,
R^dL_\perp^{d-1} A_d(z_{\textrm{max}},a)\:.
\end{equation}
In order to isolate the divergence of (\ref{strip area}) as $a \rightarrow 0$, we write the integral as follows
\begin{eqnarray}
\label{split 0}
A_d(z_{\textrm{max}},a) &=&
\int_a^{z_{\textrm{max}}} 
\frac{2}{w^d}\,dw+
\int_a^{z_{\textrm{max}}} 
\frac{2}{w^d}\left(\frac{z_{\textrm{max}}^d}{\sqrt{f(w)}\,\sqrt{z_{\textrm{max}}^{2d}-w^{2d}}}-1\right) dw
\\
\label{split 0bis}
\rule{0pt}{.7cm}
& \equiv & 
\frac{2}{(d-1)\,a^{d-1}}
+ \mathcal{A}_d(z_{\textrm{max}},a)\;.
\end{eqnarray}
The second integral in (\ref{split 0}) is finite as $a \rightarrow 0$ because we have either $f(w)=1+O(w^{d+1})$ or $f(w)=1+O(w^{d})$ for $w \rightarrow 0$ (see (\ref{charged BH metric z}) and (\ref{lifshitz metric}) respectively). In (\ref{split 0bis}) we have introduced the finite term in the UV cutoff expansion
\begin{equation}
\label{calArea}
\mathcal{A}_d(z_{\textrm{max}},a) 
\,\equiv\,
\int_a^{z_{\textrm{max}}} 
\frac{2}{w^d}\left(\frac{z_{\textrm{max}}^d}{\sqrt{f(w)}\,\sqrt{z_{\textrm{max}}^{2d}-w^{2d}}}-1\right) dw
-\frac{2}{(d-1)z_{\textrm{max}}^{d-1}}\;.
\end{equation}
In this paper we will be mainly interested $O(1)$ term in the $a$ expansion, which is $\mathcal{A}_d(z_{\textrm{max}},0)$. In the figure \ref{plotA0} this term is shown for the charged black hole in $AdS_4$ at zero temperature. As the turning point $z_{\textrm{max}}$ approaches the horizon $z_0$, it develops a divergence we are going to study.

\noindent Equivalently, we can isolate the divergence for small $a$ in the integral of (\ref{strip area}) as follows
\begin{eqnarray}
\label{split 1}
A_d(z_{\textrm{max}},a) &=&
\int_a^{z_{\textrm{max}}} 
\frac{2}{w^d \sqrt{f(w)}}\,dw
+
\int_a^{z_{\textrm{max}}} 
\frac{2}{w^d \sqrt{f(w)}}\left(\frac{z_{\textrm{max}}^d}{\sqrt{z_{\textrm{max}}^{2d}-w^{2d}}}-1\right) dw\,
\hspace{1cm}\\
\label{I def}
\rule{0pt}{.7cm}
&\equiv&
\int_a^{z_{\textrm{max}}} 
\frac{2}{w^d \sqrt{f(w)}}\,dw + I_d(a, z_{\textrm{max}})\;.
\end{eqnarray}
The finite term in the expansion for small $a$ is now given by $I_d(0, z_{\textrm{max}})$ plus a contribution from the first integral in (\ref{I def}). The distinction between (\ref{split 0}) and (\ref{split 1}) is obviously meaningless for $AdS_{d+2}$, where $f(z)=1$ identically.
The splitting (\ref{split 1}) has been used in the appendix \ref{app more on nh} to get some insights about the expansion of the finite term of the minimal area as a power series as $z_0- z_{\textrm{max}}$ and the possibility to approximate it through the near horizon geometry.

\subsection{A  more general ansatz}
\label{more general ansatz section}

In this section we consider a more complicated expression for the metric in order to include other kind of black holes in our discussion.
Let us take a $D+1$ dimensional spacetime and the following ansatz for the metric on the constant time slice
\begin{equation}
\label{ds2 general ansatz}
ds_0^2\,=\,\left(\frac{dr^2}{A(r)^2}+B(r)^2 d\vec{x}^2\right)e^{-\frac{D-1-d}{d}\chi(r)}
+R^2 e^{\chi(r)} \gamma^{\textrm{(c)}}_{ij}(r) d\theta^i \theta^j
\end{equation}
where $d\vec{x}^2$ gives the metric of $\mathbb{R}^d$ 
and $\gamma^{\textrm{(c)}}_{ij}(r)$ is the metric of a $D-1-d$ dimensional compact manifold $\mathcal{M}_{\textrm{c}}$. The boundary is at large $r$ and we assume the occurrence of a horizon at $r=r_h$. \\
Let us take a strip specified by the function $r=r(x_d)$. Then, the metric induced on it reads
\begin{equation}
ds^2_{\textrm{ind}}\,=\,\left(B(r)^2 (dx_1^2+\dots dx_{d-1}^2)
+\left[\,\frac{(r')^2}{A(r)^2}+B(r)^2\right]dx^2_d  \right)
e^{-\frac{D-1-d}{d}\chi(r)}
+R^2 e^{\chi(r)} \gamma^{\textrm{(c)}}_{ij}(r) d\theta^i \theta^j\,.
\end{equation}
To compute the area, we have to integrate $\sqrt{\textrm{det}(G_{\textrm{ind}})}$ over the strip $A$. In such determinant the dependence on $\chi(r)$ simplifies and therefore it does not occur anymore.
If $\textrm{det}(\gamma^{\textrm{(c)}})$ does not depend on $r$, then the area of the surface is given by
\begin{equation}
\label{area r}
\textrm{Area}(\gamma_A)\,=\,
\big[R^{D-1-d} \,\textrm{Vol}(\mathcal{M}_{\textrm{c}}) \big]\,
2L_\perp^{d-1}
\int_0^{L/2} B(r)^{d}\,
\sqrt{1+\frac{(r')^2}{A(r)^2 B(r)^2}}\,dx_d
\end{equation}
where $L_\perp$ is the width of the strip along the directions $x_1 , \dots , x_{d-1}$ and $\textrm{Vol}(\mathcal{M}_{\textrm{c}})$.\\
As done above, we take as Lagrangian density $\mathcal{L}_{\textrm{strip}}[r(x_d)]$ the integrand of (\ref{area r}) and compute the momentum $p_r \equiv \partial\mathcal{L}_{\textrm{strip}}/\partial r'$. Then, being $\mathcal{L}_{\textrm{strip}}$ independent of $x_d$, we can employ the conserved quantity $\mathcal{H}_{\textrm{strip}} \equiv p_r r' -\mathcal{L}_{\textrm{strip}}$.
Since at the minimum value of $r$ we have $r'(0)=0$, we set $\mathcal{H}_{\textrm{strip}}^2 \equiv B(r_{\textrm{min}})^{2d}$. 
This allows us to write (\ref{area r}) as follows
\begin{equation}
\label{area r infinity}
\textrm{Area}(\gamma_A)\,=\,
\big[R^{D-1-d} \,\textrm{Vol}(\mathcal{M}_{\textrm{c}}) \big]\,
2L_\perp^{d-1}
\int_{r_{\textrm{min}}}^{\infty} 
\frac{B(r)^{2d-1}}{A(r)\sqrt{B(r)^{2d}-B(r_{\textrm{min}})^{2d}}}\,dr\;.
\end{equation}
It is also important to express $L=L(r_{\textrm{min}})$ and it reads
\begin{equation}
\label{L ABansatz}
L\,=\,2
\int_{r_{\textrm{min}}}^{\infty} 
\frac{B(r_{\textrm{min}})^{d}}{A(r)B(r)\sqrt{B(r)^{2d}-B(r_{\textrm{min}})^{2d}}}\,dr\;.
\end{equation}
We require to have $AdS_{d+2}$ at large $r$, which means to impose
\begin{equation}
\label{AB large r}
A(r)^2\,=\,
\frac{r^2}{R^2}+O(1)
\hspace{1cm}
B(r)^2\,=\,
\frac{r^2}{R^2}+O(1)
\hspace{1cm}
\chi(r)\,\rightarrow\, 0
\hspace{1cm}
r\,\rightarrow\, + \infty\;.
\end{equation}
Because of this asymptotic behavior, the integral in (\ref{area r infinity}) is divergent. Thus, one introduces the cut off $\alpha$ at large $r$, obtaining for the regularized area
\begin{eqnarray}
\label{area r reg}
\frac{\textrm{Area}(\gamma_A)}{R^{D-1-d} \,\textrm{Vol}(\mathcal{M}_{\textrm{c}})} & &
\hspace{-.5cm}=\;
2L_\perp^{d-1}
\int_{r_{\textrm{min}}}^{\alpha} 
\frac{B(r)^{2d-1}}{A(r)\sqrt{B(r)^{2d}-B(r_{\textrm{min}})^{2d}}}\,dr \\
\label{area r finite integ 1}
\rule{0pt}{.9cm}
& &
\hspace{-3.3cm}=\;\;
\frac{2L_\perp^{d-1} }{(d-1)R^{d-2}}\,\Big(\alpha^{d-1}-r_{\textrm{min}}^{d-1}\Big)
+
2L_\perp^{d-1}  \int_{r_{\textrm{min}}}^{\alpha} 
 \left(
\frac{B(r)^{2d-1}}{A(r)\sqrt{B(r)^{2d}-B(r_{\textrm{min}})^{2d}}}
-\left(\frac{r}{R}\right)^{d-2}
\,\right)dr
\nonumber\\
& &
\end{eqnarray}
where the integral in (\ref{area r finite integ 1}) is finite when $\alpha \rightarrow \infty$, once the asymptotic behavior (\ref{AB large r}) has been assumed. At this point, the finite term of area integral we are interested in is given by the sum of the integral and of the term proportional to $r_{\textrm{min}}^{d-1}$ in (\ref{area r finite integ 1}).
We remark that (\ref{L(zmax)}) and (\ref{strip area}) are special cases of (\ref{L ABansatz}) and (\ref{area r reg}) respectively. Indeed they are recovered by choosing
\begin{equation}
A(r)\,=\,
\frac{r}{R}\,\sqrt{f(r)}
\hspace{2.2cm}
B(r)\,=\, \frac{r}{R}
\end{equation}
and adopting the variable $z \equiv R^2/r$. 
The formula for the holographic entanglement entropy then gives
\begin{equation}
S_A\,=\,
\frac{\textrm{Area}(\gamma_A)}{4 G_N^{(D+1)}}\,=\,
\frac{2L_\perp^{d-1}}{4 G_N^{(d+2)}}
\int_{r_{\textrm{min}}}^{\alpha} 
\frac{B(r)^{2d-1}}{A(r)\sqrt{B(r)^{2d}-B(r_{\textrm{min}})^{2d}}}\,dr
\end{equation}
where we have used that $G_N^{(D+1)}=G_N^{(d+2)}[R^{D-1-d} \,\textrm{Vol}(\mathcal{M}_{\textrm{c}})]$.
Notice that the compact part enters through Kaluza-Klein reduction in the Newton's constant also in this case where a warping factor occurs between the compact and the non compact part \cite{Barbon:2008sr, Barbon:2008ut}.


\section{Expansion of the finite term near the horizon}
\label{section near horizon}

In this section we study the finite term of the holographic entanglement entropy introduced in the previous section. In particular, we consider the leading term of its expansion as the turning point of the minimal surface approaches the horizon, which means that the width $L$ of the strip in the boundary becomes large. As examples, we analyze the charged black hole in $AdS_{d+2}$ (section \ref{section nh charged bh}), the warped black hole of \cite{Herzog:2009gd} (section \ref{section warped bh}) and the perturbation of the Lifshitz background found in \cite{Gubser:2009cg} within the context of the Abelian Higgs model of \cite{Gubser:2008px} (section \ref{section perturbed lifshitz}).\\
The finite term in the expansion for small UV cutoff $a$ is given by $\mathcal{A}_d(z_{\textrm{max}},0)$, defined in (\ref{calArea}). 
In order to consider its expansion as the turning point $z_{\textrm{max}}$ of the minimal surface gets close to the horizon, we take
\begin{equation}
\label{eps def}
z_{\textrm{max}}  \equiv z_0- \varepsilon \zeta_{\textrm{max}} 
\hspace{1.5cm}
\varepsilon \,\rightarrow\,0
\hspace{1.5cm}
\textrm{finite $\zeta_{\textrm{max}}$}
\end{equation}
and change the integration variable in (\ref{calArea}) according to this expansion, i.e. we set $w = z_0- \varepsilon \zeta$, where $0 < \zeta_{\textrm{max}} < \zeta$. Then, the finite term can be written as follows
\begin{equation}
\label{A0 expanded}
\mathcal{A}_d(z_{\textrm{max}},0)\,=\,
\sum_{k \in B }
\varepsilon^{k}
 \int_{\frac{z_0}{\varepsilon}}^{\zeta_{\textrm{max}}} 
\mathcal{I}_{k}(\zeta,\zeta_{\textrm{max}}) d\zeta
-\frac{2}{(d-1)z_{\textrm{max}}^{d-1}}
\end{equation}
where $B \subset [k_{\textrm{min}}, \infty) \subset \mathbb{Q}$ is some discrete set of increasing rational numbers, which are not necessarily positive ($k_{\textrm{min}} <0$).
For instance, in the case of the charged black hole with $d=2$ we have $k \in \{ -1/2,1/2,1,3/2,\dots\}$.
In order to write $\mathcal{A}_d(z_{\textrm{max}},0)$ as an expansion in terms of powers of $z_0- z_{\textrm{max}}$, we have to compute the definite integrals occurring at each $k$ and then expand each of them for small $\varepsilon$. Then this expansion can be written in powers of 
$z_0- z_{\textrm{max}}$ by using the definition $\varepsilon=(z_0- z_{\textrm{max}})/\zeta_{\textrm{max}}$ from (\ref{eps def}).\\
In all examples we have considered we find that this method provides only the divergent term as  $z_{\textrm{max}} \rightarrow z_0$. This is due to the fact that all the integrals occurring in (\ref{A0 expanded}) give a contribution to the finite term of the expansion. \\
The same procedure just described to expand the integral $\mathcal{A}_d(z_{\textrm{max}},0)$ can be applied to the integral (\ref{L(zmax)}) as well, obtaining $L$ as an expansion in powers of $z_0- z_{\textrm{max}}$. It is then useful to compare the divergences of these two quantities as $z_{\textrm{max}} \rightarrow z_0$ in order to see how the finite term of the entanglement entropy scales with the width of the strip, and therefore with the volume.

\subsection{Charged black hole}
\label{section nh charged bh}

In this section we apply the method just described to the charged black hole in $AdS_{d+2}$ in its three different regimes of neutrality, extremality and non extremality. The metric and its properties are reviewed in the appendix \ref{app chargedBH}.\\
The metric of the charged black hole in $AdS_{d+2}$ reads
\begin{equation}
\label{charged BH metric z}
\frac{ds^2}{R^2}\,=\,\frac{-f dt^2+ d\vec{x}^2}{z^2}+\frac{dz^2}{f z^2}
\hspace{1.6cm}
f\,=\,1+Q^2\left(\frac{z}{R^2}\right)^{2d}-M\left(\frac{z}{R^2}\right)^{d+1}
\end{equation}
where $M$ is the mass and $Q$ is the charge of the black hole. The radial direction is parameterized by $z$ and the boundary is at $z=0$. 
The position $z_0$ of the horizon is given by the smallest zero of the emblacking function $f(z)$. Since the metric (\ref{charged BH metric z}) falls into the class of metrics described by (\ref{ds20 ansatz}), we can employ the formulas discussed in the section \ref{section strip BH f(z)}.\\
$\bullet$ {\bf Schwarzschild black hole.}
As a first example, we consider the Schwarzschild black hole, which is given by (\ref{charged BH metric z}) with $Q=0$.
By performing the expansion described above, we find
\begin{equation}
L\,=\,
-\,\frac{\sqrt{2}\,z_0}{\sqrt{d(d+1)}}\,\log(z_0-z_{\textrm{max}})
+ O(1)
\end{equation}
and
\begin{equation}
\label{expansion nh shwarz}
\mathcal{A}_d(z_{\textrm{max}},0)
\,=\,
-\,\frac{\sqrt{2}}{\sqrt{d(d+1)}\,z_0^{d-1}}\,\log(z_0-z_{\textrm{max}})
+ O(1)
\,=\,
\frac{L}{z_0^{d}}
+ O(1)
\end{equation}
where we recall that the horizon $z_0$ is related to the temperature as $T=(d+1)/(4\pi z_0)$.
The case $d=3$ was considered in \cite{Ryu:2006ef}.

\begin{figure}[h]
\begin{tabular}{ccc}
\includegraphics[width=7.5cm]{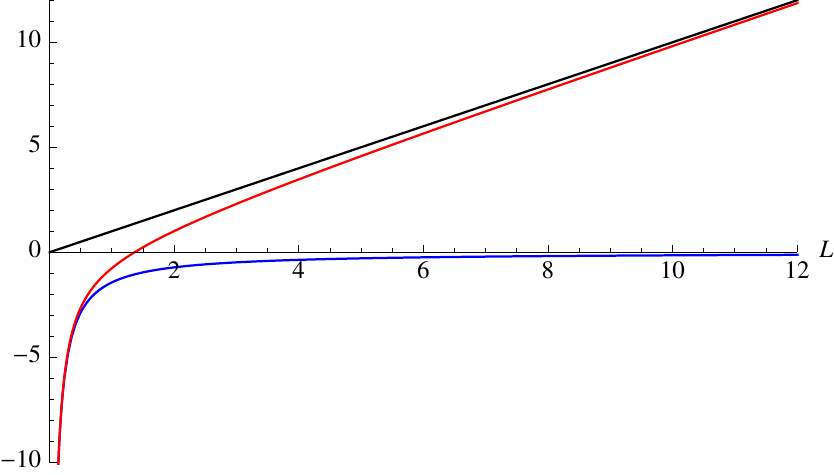}
& & 
\includegraphics[width=7.5cm]{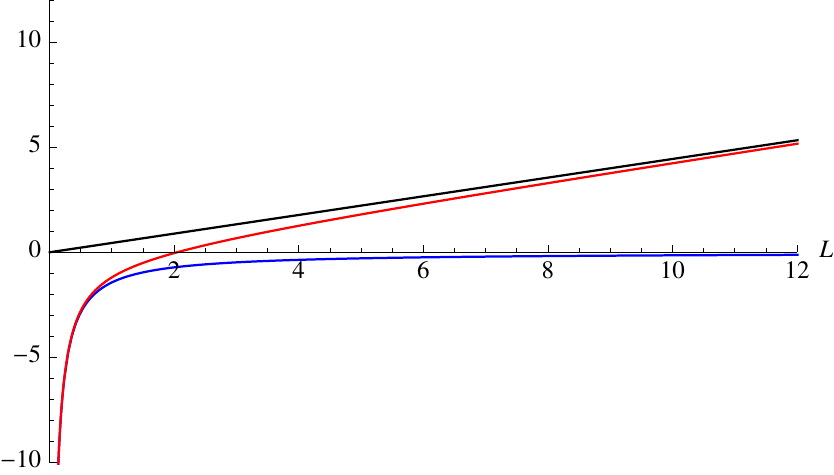}
\end{tabular}
\caption{Extremal charged black hole in $AdS_4$ with $z_0=1$ (left) and $z_0=1.5$ (right). 
Plot of the finite term $\mathcal{A}_2(z_{\textrm{max}},0)$ as a function of $L$ (red line). For small $L$ it recovers the corresponding quantity for $AdS_4$ 
(blue curve) obtained from (\ref{Atilde free}). The black line provides the large $L$ behavior given by (\ref{expansion nh rn T0}).
\label{finite area L}}
\end{figure}

\noindent 
$\bullet$ {\bf Extremal charged black hole.}
When $Q\neq 0$ and $T=0$ this analysis leads to
\begin{equation}
L\,=\,
\frac{\sqrt{2}\,\pi\,z_0^{3/2}}{d \sqrt{d+1}\,\sqrt{z_0-z_{\textrm{max}}}}
+ O(1)
\end{equation}
and
\begin{equation}
\label{expansion nh rn T0}
\mathcal{A}_d(z_{\textrm{max}},0)
\,=\,
\frac{\sqrt{2}\,\pi\,z_0^{3/2-d}}{d \sqrt{d+1}\,\sqrt{z_0-z_{\textrm{max}}}}
+ O(1)
\,=\,
\frac{L}{z_0^{d}}
+ O(1)\;.
\end{equation}
In the figure \ref{finite area L} (see \cite{Barbon:2008sr}) we show $\mathcal{A}_d(z_{\textrm{max}},0)$ in terms of $L$ for the extremal case. When $L$ is small we are close to the boundary and the curve reproduces the one of $AdS_4$, as expected. By comparing the two plots in the figure, one can check the dependence on $z_0$ in (\ref{expansion nh rn T0}).

\noindent 
$\bullet$  {\bf Non extremal charged black hole.}
The same method applied for $T>0$ leads to
\begin{equation}
\label{L exp extremal ddim}
L\,=\,
-\,\frac{\sqrt{z_0}}{\sqrt{2\pi d\, T}}\,\log(z_0-z_{\textrm{max}})
+ O(1)
\end{equation}
and
\begin{equation}
\label{expansion nh rn T}
\mathcal{A}_d(z_{\textrm{max}},0)
\,=\,
-\,\frac{\sqrt{z_0}}{z_0^d\sqrt{2\pi d\, T}}\,\log(z_0-z_{\textrm{max}})
+ O(1)
\,=\,\frac{L}{z_0^{d}}
+ O(1)\;.
\end{equation}
Comparing these three regimes of the same black hole, one learns that the finite term of the holographic entanglement entropy diverges like the width $L$ (and therefore like the volume) of the strip in the boundary. The distinguished feature is the kind of divergence of $\mathcal{A}_d(z_{\textrm{max}},0)$ and $L$ as $z_{\textrm{max}} \rightarrow z_0$. This is determined by the near horizon geometry which is given by $f(z)=O(z_0-z)$ for the Schwarzschild and the non extremal case and by $f(z)=O\big((z_0-z)^2\big)$ for the extremal case (see the appendix \ref{app chargedBH}). As a check, one can perform the expansion of the finite term just described substituting to the emblacking function $f(z)$ its near horizon approximation $f_{\textrm{nh}}(z)$ and verify that the same divergence shown above are obtained.\\
We remark that for all the black holes we are considering the horizon is non compact; therefore the wrapping of the minimal surface around the horizon in the large $L$ limit described e.g. in \cite{Ryu:2006bv, Ryu:2006ef, Headrick:2007km, Azeyanagi:2007bj} does not occur.
In the appendix \ref{app more on nh} we employ the splitting (\ref{split 1}) of the finite term to study the $O(1)$ term in (\ref{expansion nh rn T0}) and discuss the approximation obtained by using the near horizon geometry.

\subsection{Warped black hole}
\label{section warped bh}

In this section we employ the observation just made about the role of the region close to the horizon and apply the expansion described in (\ref{eps def}) and (\ref{A0 expanded}) to a black hole where only the near horizon geometry is known.\\
In \cite{Herzog:2009gd} a minimal consistent truncation of the type IIB supergravity has been considered by the following metric
\begin{equation}
\label{warped bh metric}
ds^2\,=\,e^{-\frac{5}{3}\chi} ds^2_M +R^2 e^\chi
\left[\,\frac{e^{-4\eta}}{9}\bigg( d\psi+\sum_{i=1}^2 \cos\phi_i\bigg)^2
+\frac{e^{\eta}}{6} \sum_{i=1}^2 \Big(d\theta_i^2+\sin^2\theta_i\,d\phi^2_i\Big)
\,\right]
\end{equation}
where the non compact space $M$ is given by
\begin{equation}
ds^2_M \,=\,-\,g e^{-w} dt^2+\frac{dr^2}{g}+\frac{r^2}{R^2} \sum_{i=1}^3 dx_i^2\;.
\end{equation}
The functions $\chi$, $\eta$, $g$ and $w$ depend on the coordinate $r$ only. 
The geometry (\ref{warped bh metric}) is required to provide $AdS_5 \times T^{1,1}$ on the boundary, i.e. at large $r$.\\
In \cite{Herzog:2009gd} the equations of motion coming from the effective Lagrangian have been solved numerically; nevertheless analytic formulae have been found in some limits. We are interested in the $T=0$ regime, for which the first term of a series expansion near the horizon is given. The novel feature is that the near horizon region is a warped product $AdS_2 \times \mathbb{R}^3 \times T^{1,1}$.
As discussed in \cite{Herzog:2009gd}, one can employ the symmetries of the problem to set to one both the $AdS$ radius and the position of the horizon, but we prefer to keep $r_0$ generic for clearness.\\
The metric (\ref{warped bh metric}) falls into the general class considered in the section \ref{more general ansatz section} through the ansatz (\ref{ds2 general ansatz}) by choosing $D=9$, $d=3$ and
\begin{equation}
A(r)^2\,=\,g(r)
\hspace{2cm}
B(r)^2\,=\,\frac{r^2}{R^2}\;.
\end{equation}
The analytic behavior near the horizon in the $T=0$ case reads \cite{Herzog:2009gd}
\begin{equation}
\label{klebanov bh g}
g(r) \,=\, b (r-r_0)^{13/3} + \dots
\hspace{2cm} b\,\equiv\,\frac{93312\,\sqrt[3]{12}}{25}\;.
\end{equation}
As checked in the section \ref{section nh charged bh} for the charged black hole, the near horizon region determines the leading divergence of the finite term of the holographic entanglement entropy as the strip in the boundary becomes large.
Thus, we perform the expansion discussed at the beginning of the section \ref{section near horizon} by using the near horizon geometry (\ref{klebanov bh g}) instead of the full metric (which is still analytically unknown).
Introducing $r_{\textrm{min}} = r_0+ \varepsilon\rho_{\textrm{min}}$ with finite $\rho_{\textrm{min}}$ and changing the integration variable accordingly ($r = r_0+ \varepsilon\rho$), we get for the leading behavior of the integral in (\ref{L ABansatz}) the following result
\begin{equation}
\label{L warped bh}
\frac{L}{2} \,=\,
\frac{1}{\sqrt{6b \,r_0}\,\varepsilon^{5/3}}\,
\int^{\infty}_{\rho_{\textrm{min}}}\frac{d\rho}{\rho^{13/6}(\rho-\rho_{\textrm{min}})}\,+ \dots
\,=\,
\frac{1}{\sqrt{6b \,r_0}}
\frac{\sqrt{\pi}\,\Gamma\big(\tfrac{5}{3}\big)}{\Gamma\big(\tfrac{13}{6}\big)\,(r_{\textrm{min}}-r_0)^{5/3}}
 +\,\dots
\end{equation}
where $\dots$ denote higher orders in $\varepsilon$.
The same procedure can be applied to the integral in (\ref{area r finite integ 1}) which provides the leading divergence of the finite term in the holographic entanglement entropy as $r_{\textrm{min}}$ approaches the horizon. The result reads
\begin{equation}
 \int_{r_{\textrm{min}}}^{\infty} 
 \Bigg(\,
\frac{r^5}{\sqrt{g(r)}\,\sqrt{r^{6}-r_{\textrm{min}}^{6}}}
-r \Bigg)dr\,=\,
\frac{r_0^{5/2}}{\sqrt{6b}}
\frac{\sqrt{\pi}\,\Gamma\big(\tfrac{5}{3}\big)}{\Gamma\big(\tfrac{13}{6}\big)\,(r_{\textrm{min}} -r_0)^{5/3}}
 +\,\dots
\,=\,  r_0^3 \, L  +\,\dots
\end{equation}
where in the last step we have used (\ref{L warped bh}).
Thus, also in this case the expected behavior for $r_{\textrm{min}} \rightarrow r_0$ is recovered (here we have $d=3$).

\subsection{Perturbed Lifshitz background}
\label{section perturbed lifshitz}

The Lifshitz background is defined by a metric which is scale invariant if the space coordinates and the time coordinate scale with a different power. The relative scale dimension of time and space is the dynamical exponent. This parameter usually occurs in the time component of the metric; therefore it does not affect the computation of the holographic entanglement entropy, which involves the metric on a constant time slice. An example of this type is considered in the section \ref{section lifshitz bh}. Instead, when the dynamical exponent occurs in some spatial component of the metric, then it usually turns out to be involved non trivially in the holographic computation of the entanglement entropy \cite{Azeyanagi:2009pr}. In this section we consider an example of this type. \\
A perturbation of the Lifshitz background through a formal parameter expansion was studied in \cite{Gubser:2009cg} as a solution of the Abelian Higgs model in $AdS_4$ \cite{Gubser:2008px}, introduced to describe superconducting black holes. The metric to consider reads
\begin{equation}
\label{metric perturbed lifshitz}
ds^2\,=\,
-\,g(r)^2 dt^2+\frac{r^2}{R^2}\,d\vec{x}^2+e^{2b(r)} \frac{R^2}{r^2}\,dr^2\;.
\end{equation}
In \cite{Gubser:2009cg}  it was found that the Lifshitz background is a solution and also its perturbation of the following form is allowed
\begin{equation}
\label{g and b gubser}
g(r)\,=\,\left(\frac{r}{R}\right)^\omega+\lambda \,g_1(r) + O(\lambda^2)
\hspace{1.5cm}
b(r)\,=\,\lambda\,c\,r^\gamma+ O(\lambda^2)
\end{equation}
where $\lambda$ is a formal expansion parameter and $\gamma$ depends on the dynamical exponent $\omega$ besides other parameters of the model. The explicit expression of $\gamma$ is not important for our discussion.
Notice that the dynamical exponent affects the spatial part of the metric through the perturbation of the Lifshitz background, and therefore it occurs in the computation of the holographic entanglement entropy. Since (\ref{metric perturbed lifshitz}) on a constant time slice is a special case of the ansatz considered in the section \ref{more general ansatz section}, we can employ the results discussed there.
From (\ref{area r finite integ 1}) with $d=2$, $B(r)=r/R$ and $A(r)=e^{-b(r)} r/R$, we find that the finite term in the holographic entanglement entropy is provided by the following integral
\begin{eqnarray}
\label{lif integral hee}
 \int_{r_{\textrm{min}}}^{\alpha} 
 \left(
\frac{e^{b(r)}}{\sqrt{1-(r_{\textrm{min}}/r)^4}}-1\right)
dr
&=&
 \int_{r_{\textrm{min}}}^{\alpha} 
 \left(
\frac{1}{\sqrt{1-(r_{\textrm{min}}/r)^4}}-1\right)
dr\\
\rule{0pt}{.7cm}
& &\hspace{.5cm}
+ \;\lambda\, c
 \int_{r_{\textrm{min}}}^{\alpha} 
\frac{r^\gamma}{\sqrt{1-(r_{\textrm{min}}/r)^4}}
\,dr
+O(\lambda^2)\;.
\nonumber
\end{eqnarray}
We are mainly interested in the $O(\lambda)$ term in the r.h.s. of (\ref{lif integral hee})  because the $O(1)$ one provides the result of $AdS_4$ and of the Lifshitz background in four dimensions (they have the same entanglement entropy because their metric differs only in the time component).
In (\ref{lif integral hee}) we cannot go to $O(\lambda^2)$ because it involves the $O(\lambda^2)$ of $b(r)$ in (\ref{g and b gubser}), which is not known; but the $O(\lambda)$ term is already interesting because it contains the dynamical exponent through $\gamma$. To get  a finite result from the integral at $O(\lambda)$ in (\ref{lif integral hee}) when $\alpha \rightarrow \infty$ we need $\gamma<-1$. Then we have
\begin{equation}
\label{pert lifshitz area 2}
 \int_{r_{\textrm{min}}}^{\alpha} 
\frac{r^\gamma}{\sqrt{1-(r_{\textrm{min}}/r)^4}}\,dr
\,=\,
\frac{r_{\textrm{min}}^{\gamma+1}}{4}
\, B_{\rho^4}\Big(-\frac{1+\gamma}{4},\frac{1}{2}\Big)
\Big|_{\frac{r_{\textrm{min}}}{\alpha}}^{1} 
\,=\,
\frac{\sqrt{\pi}\,\Gamma(\tfrac{-1-\gamma}{4})}{4\,\Gamma(\tfrac{1-\gamma}{4})}
\,r_{\textrm{min}}^{\gamma+1}
+O(\alpha^{1+\gamma})
\end{equation}
where we found it useful to employ the integration variable $\rho \equiv r_{\textrm{min}}/r$ and the final result is expressed in terms of the incomplete beta function $B_z(a,b)$, which reduces to the beta function $B(a,b)=\Gamma(a)\Gamma(b)/\Gamma(a+b)$ for $z=1$ and it is related to the hypergeometric function for a general $z$ as $B_z(p,q)= (z^p/p) \, _2F_1(p,1-q;1+p;z)$.\\
As for the length $L$ of the interval in the boundary, it is related to $r_{\textrm{min}}$  through the integral (\ref{L ABansatz}), which in this case can be expanded up to $O(\lambda)$, similarly to what we have done in (\ref{lif integral hee})  for the area of the minimal surface. The result is
\begin{equation}
\label{L lifshitz step1}
L \,=\, 2R^2 r_{\textrm{min}}^2
 \int_{r_{\textrm{min}}}^{\infty} 
\frac{e^{b(r)}}{r^4\,\sqrt{1-(r_{\textrm{min}}/r)^4}}\, dr
\,=\,
\frac{2R^2\,\sqrt{\pi}\,\Gamma(\tfrac{3}{4})}{r_{\textrm{min}}\,\Gamma(\tfrac{1}{4})}
+\lambda\,
\frac{cR^2\,\sqrt{\pi}\,\Gamma(\tfrac{3-\gamma}{4})}{2r_{\textrm{min}}^{1-\gamma}\Gamma(\tfrac{5-\gamma}{4})}
+O(\lambda^2)\;.
\end{equation}
Again, the first term in (\ref{L lifshitz step1}) provides the result for $AdS_4$ (see (\ref{L free})).
We can invert (\ref{L lifshitz step1}) perturbatively and find $r_{\textrm{min}}(L)$ up to $O(\lambda^2)$ terms by using that
\begin{equation}
L\,=\,c_0\,r_{\textrm{min}}^{d_0}\big[1+c_1r_{\textrm{min}}^{d_1}\lambda+O(\lambda^2)\big]
\hspace{1cm}
r_{\textrm{min}}\,=\,
\left(\frac{L}{c_0}\right)^{\frac{1}{d_0}}
\bigg[\,1-\frac{c_1}{d_0}\left(\frac{L}{c_0}\right)^{\frac{d_1}{d_0}}\lambda+O(\lambda^2)\bigg]\;.
\hspace{.5cm}
\end{equation}
In our case we find
\begin{equation}
r_{\textrm{min}}\,=\,
\frac{2R^2\,\sqrt{\pi}\,\Gamma(\tfrac{3}{4})}{\Gamma(\tfrac{1}{4})\,L}
\left[\,1+
\lambda\,
\frac{c\,\Gamma(\tfrac{3-\gamma}{4})\,\Gamma(\tfrac{1}{4})}{4\,\Gamma(\tfrac{5-\gamma}{4})\,\Gamma(\tfrac{3}{4})}\left(\frac{2R^2\,\sqrt{\pi}\,\Gamma(\tfrac{3}{4})}{\Gamma(\tfrac{1}{4})\,L}\right)^\gamma
+O(\lambda^2)
\right]\;.
\end{equation}
Plugging this result into (\ref{pert lifshitz area 2}) we find that the correction $O(\lambda)$ to the holographic entanglement entropy is proportional to
\begin{equation}
\lambda\,c \int_{r_{\textrm{min}}}^{\alpha} 
\frac{r^\gamma}{\sqrt{1-(r_{\textrm{min}}/r)^4}}\,dr
\,=\,\frac{\lambda \, c}{L^{1+\gamma}}\,
\frac{\sqrt{\pi}\,\Gamma(\tfrac{-1-\gamma}{4})}{4\,\Gamma(\tfrac{1-\gamma}{4})}
\left(\frac{2R^2\,\sqrt{\pi}\,\Gamma(\tfrac{3}{4})}{\Gamma(\tfrac{1}{4})}\right)^{\gamma+1}
\Big[1+O(\alpha^{1+\gamma})\Big]
+O(\lambda^2)\,.
\end{equation}
Since we are assuming $1+\gamma <0$ this term diverges like $L^{-(1+\gamma)}$. The interesting feature is that the dynamical exponent occurs in a non trivial way in the scaling of the finite term of the holographic entanglement entropy in terms of the width $L$ of the strip. This computation is not conclusive because it involves only the first term of a perturbative expansion, but we expect the occurrence of the dynamical exponent in such scaling also for the result computed with the full (non perturbative) expression of the metric.

\section{A Lifshitz black hole in four dimensions}
\label{section lifshitz bh}

In this section we consider the Lifshitz black hole in four dimensions ($d=2$) found in \cite{Balasubramanian:2009rx}. Because of the simple emblacking function  characterizing this black hole, we can compute the holographic entanglement entropy analytically to all order in the UV cutoff. This allows us also to check the method employed in the section \ref{section near horizon} to find the divergent term in the finite integral of the area as $z_{\textrm{max}}$ goes to the horizon $z_0$.\\
The Lifshitz black hole of \cite{Balasubramanian:2009rx} is a solution e.g. of a model in four dimensions which includes, besides gravity, a massive $U(1)$ gauge field and a strongly coupled scalar, namely a scalar without kinetic term.
Its metric reads
\begin{equation}
\label{lifshitz metric}
ds^2\,=\,-f(z) \frac{dt^2}{z^{2 \omega}}+
\frac{d\vec{x}^2}{z^2}+\frac{dz^2}{z^2 f(z)}
\hspace{2cm}
f(z) \,=\,1-\frac{z^2}{z_0^2}\;.
\end{equation}
The boundary is at $z=0$ and the range of the holographic coordinate is $(0,z_0)$. The dynamical exponent is $\omega=2$ and the bulk curvature radius $R$ has been set to one.
Near the boundary the metric (\ref{lifshitz metric}) asymptotes the Lifshitz spacetime in four dimensions with dynamical exponent equal to two.
Near the horizon the emblacking function $f(z)$ vanishes linearly and the metric on the constant time slice is (\ref{ds20 ansatz}) with the $f(z)$ given in (\ref{lifshitz metric}).\\
We remark that, since the anisotropy $\omega$ does not occur in the metric on the constant $t$ slice, we do not see the effects described in   \cite{Azeyanagi:2009pr}. In that case they have an anisotropy between two spatial directions; therefore the holographic entanglement entropy is sensible to the difference between them.\\
As first step we study the leading order for $z_{\textrm{max}} \rightarrow z_0$ of the finite term (\ref{calArea}) by employing the expansion described in the section \ref{section near horizon}. The result is
\begin{equation}
\label{lifshitz area0 leading}
\mathcal{A}_{2}(z_{\textrm{max}},0)\,=\,
- \frac{1}{\sqrt{2}\,z_0}\log(z_0-z_{\textrm{max}})
+ O(1)\;.
\end{equation}
Like in all the cases considered in the section \ref{section near horizon}, we cannot say anything about the finite term with this method.\\
For the Lifshitz black hole (\ref{lifshitz metric}) we can compute the integral in (\ref{strip area}) analytically (we find it  convenient to adopt $\tilde{z}\equiv w^2/z_{\textrm{max}}^2$ as integration variable). The result reads
\begin{equation}
\label{HEE lifshitz}
A_{2}(z_{\textrm{max}},a)\,=\,\int_a^{z_{\textrm{max}}} 
\frac{2z_{\textrm{max}}^2}{w^2 \sqrt{f_L(w)}\,\sqrt{z_{\textrm{max}}^{4}-w^{4}}}\,dw
\,=\,-\,\frac{1}{z_{\textrm{max}}}\,
\mathcal{I}\bigg(\frac{a^2}{z_{\textrm{max}}^2}\bigg)
\end{equation}
where
\begin{eqnarray}
\mathcal{I}(x)
& \equiv &
2\sqrt{1+\beta}\;
E\left(\arcsin\left(\sqrt{\frac{(1+\beta)(1-x)}{2(1-\beta x)}}\,\right)\bigg|\frac{2}{1+\beta}\right)
\\
\rule{0pt}{.9cm}& &\hspace{3.8cm}
-\;\sqrt{2}\,\beta
\;F\left(\arcsin\left(\sqrt{\frac{1-x}{1-\beta x}}\,\right)\bigg|\frac{1+\beta}{2}\right)
-2\,\sqrt{\frac{1-x^2}{x(1-\beta x)}}
\nonumber 
\end{eqnarray}
being $\beta\equiv (z_{\textrm{max}}/ z_0)^2$ and the function $F(x|m)$ and $E(x|m)$ the incomplete elliptic integrals of the first and of the second kind respectively. Notice that the upper extremum of integration in (\ref{HEE lifshitz}) gives a vanishing contribution.
Expanding (\ref{HEE lifshitz}) for small UV cutoff $a$ we find
\begin{eqnarray}
A_{2}(z_{\textrm{max}},a) &=&
\frac{2}{a}-\frac{f_0(\beta)}{z_{\textrm{max}}}
-\frac{\beta\,a}{z_{\textrm{max}}^2}
-\left(\frac{5}{3}-\frac{13}{12}\,\beta^2\right)\frac{a^3}{z_{\textrm{max}}^4}
- \left(\frac{13}{10}\beta -\frac{43}{40}\,\beta^3\right)\frac{a^5}{z_{\textrm{max}}^6}
\hspace{.4cm}
\\
\rule{0pt}{.7cm}& &\hspace{6cm}
-\left(\frac{11}{28}+\frac{47}{56}\,\beta^2-\frac{445}{448}\,\beta^4\right)\frac{a^7}{z_{\textrm{max}}^8}
+O(a^9)\nonumber
\end{eqnarray}
with the function $f_0(\beta)$ occurring in the finite term of this expansion given by
\begin{equation}
f_0(\beta) \,\equiv\,
2\sqrt{1+\beta}\,
E\left(\arcsin\left(\sqrt{\frac{1+\beta}{2}}\,\right)\bigg|\frac{2}{1+\beta}\right)
-\sqrt{2}\,\beta \,
K\left(\frac{1+\beta}{2}\right)
\end{equation}
where $K(z)$ is the complete elliptic integral of the first kind. As $z_{\textrm{max}} \rightarrow z_0$ we get 
\begin{equation}
\label{lifshitz finite term exp}
-\frac{f_0(\beta)}{z_{\textrm{max}}}\,=\,
- \frac{1}{\sqrt{2}\,z_0}\log(z_0-z_{\textrm{max}})
+ O(1)
\end{equation}
which confirms the result (\ref{lifshitz area0 leading}) found through the method described in the section \ref{section near horizon}.\\
For this black hole we can compute also the integral (\ref{L(zmax)}) as done for the one in (\ref{HEE lifshitz}). Again, the upper extremum of the definite integral  gives a vanishing contribution. The result reads
\begin{equation}
L = \frac{2 z_{\textrm{max}}}{\beta \sqrt{1+\beta}}
\left[
\,F\left(\arcsin\left(\frac{\sqrt{1+\beta}}{\sqrt{2}}\,\right)\bigg|\frac{2}{1+\beta}\right)
-(1-\beta)\,
\Pi\left(\frac{2\beta}{1+\beta} \, ; \arcsin\left(\frac{\sqrt{1+\beta}}{\sqrt{2}}\,\right)\bigg|\frac{2}{1+\beta}\right)
\right]
\end{equation}
where $\Pi(x,\phi | m)$ is the incomplete elliptic integral of the third kind.
When $z_{\textrm{max}} \rightarrow z_0$ we have
\begin{equation}
L\,=\,
- \frac{z_0}{\sqrt{2}}\log(z_0-z_{\textrm{max}})
+ O(1)\;.
\end{equation}
Combining this result with (\ref{lifshitz finite term exp}) we obtain
\begin{equation}
\mathcal{A}_2(z_{\textrm{max}}),0)\,=\, \frac{L}{z_0^2} + O(1)
\end{equation}
as expected. Besides providing another check for the method discussed in the section \ref{section near horizon}, this is the first case of a black hole whose holographic entanglement entropy can be computed analytically. 


\section{Two disconnected strips}
\label{section 2 strips}

In this section we consider the case of a region $A$ in the boundary made by two parallel strips.  
In particular, following \cite{Headrick:2010zt}, we study the transition of the mutual information in $AdS_{d+2}$ (section \ref{AdS 2intervals}) and in the charge black hole background (section \ref{section 2 int charged bh}).\\
Let us consider a spatial slice of the boundary theory with two parallel strips $A_1$ and $A_2$ whose widths are $L_1$ and $L_2$ respectively and separated by a distance $L_0$. 
As recalled in the introduction, the natural quantity to study for two disconnected regions is the mutual information $M_A \equiv S_{A_1}+S_{A_2}-S_{A_1 \cup A_2}$ because it is UV finite.\\
In order to find the minimal surface associated to the region $A= A_1 \cup A_2$ in the holographic computation, we have to consider two pairs of disjoint surfaces extended in the bulk whose boundary coincides with the boundary of the two strips. 
Together with the region $A$, the first pair of surfaces encloses a connected volume of the bulk, while the second one encloses two disconnected volumes of the bulk. 
The strong subadditivity inequalities guarantee that the pair of intersecting surfaces in the bulk whose boundary coincides with $\partial A$ is not minimal \cite{Headrick:2007km, Hubeny:2007re, Headrick:2010zt}.
The divergent term giving the area law is the same for both these two pairs of surfaces because they share the same boundary. Thus, in order to find the pair with minimal surface, we have to consider the finite term (in the UV cutoff) of the integrals giving the area of the pair of surfaces.\\
We find it useful here to change slightly the notation for the finite part (\ref{calArea}) of the holographic entanglement entropy by introducing $\tilde{A}_d(L)\equiv\mathcal{A}_d(z_{\textrm{max}},0)$ where $z_{\textrm{max}}= z_{\textrm{max}}(L)$ is the inverse function of (\ref{L(zmax)}). 
Thus, we consider
\begin{equation}
\label{Sd def}
S_d(L_1,L_2;L_0) \,\equiv\, 
\textrm{min}\big[\underbrace{\tilde{A}_d(L_1)+\tilde{A}_d(L_2)}_{\textrm{disconnected volumes}}\,;
\underbrace{\tilde{A}_d(L_0)+\tilde{A}_d(L_1+L_0+L_2)}_{\textrm{connected volume}} \big]
\end{equation}
which occurs in the mutual information for the finite parts
\begin{equation}
\label{MI def}
M_d(L_1,L_2;L_0) \,\equiv\, \tilde{A}_d(L_1)+\tilde{A}_d(L_2) 
- S_d(L_1,L_2;L_0)\;.
\end{equation}
We remark that in (\ref{MI def}) we talk about mutual information with a slight abuse of notation because the mutual information is given by (\ref{MI def}) multiplied by a factor $R^d L_\perp^{d-1}/(4 G_N^{(d+2)})$ coming from (\ref{RT prescription}) and (\ref{strip area}). We made this choice for clearness and we believe it will not mislead the reader.\\
The mutual information (\ref{MI def}) is zero when the minimal surface is given by the pair of surfaces enclosing the disconnected volumes and it is positive when the minimal surface corresponds to the pair of surfaces enclosing the is the connected volume. The transition of the mutual information (\ref{MI def}) from zero to a positive value occurs when the two terms compared in (\ref{Sd def}) are equal, i.e.
\begin{equation}
\label{transition eq}
\tilde{A}_d(L_1)+\tilde{A}_d(L_2) = 
\tilde{A}_d(L_0)+\tilde{A}_d(L_1+L_0+L_2)\;.
\end{equation}
In the remaining part of this section we study this equation in the special case of equal strips, namely $L_1= L_2$. First we consider $AdS_{d+2}$, where some analytic result can be found, and then the charged black hole in $AdS_{d+2}$.
\begin{figure}[h]
\begin{tabular}{ccc}
\includegraphics[width=7.5cm]{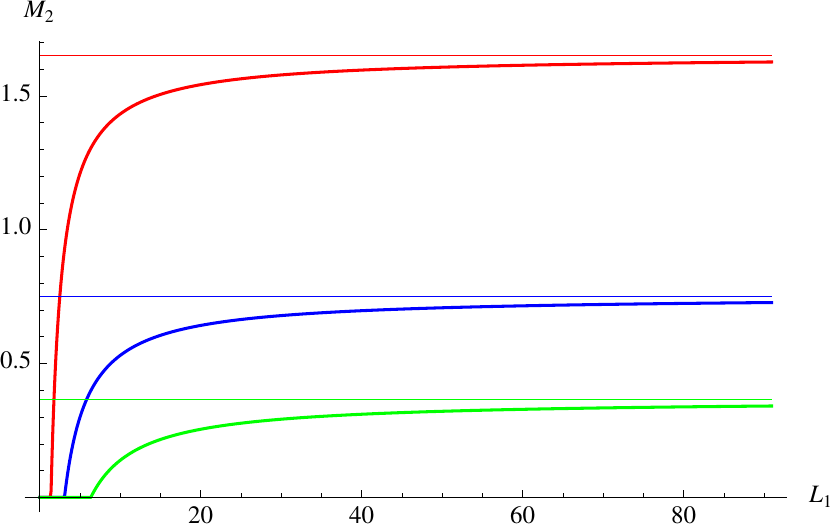}
& & 
\includegraphics[width=7.5cm]{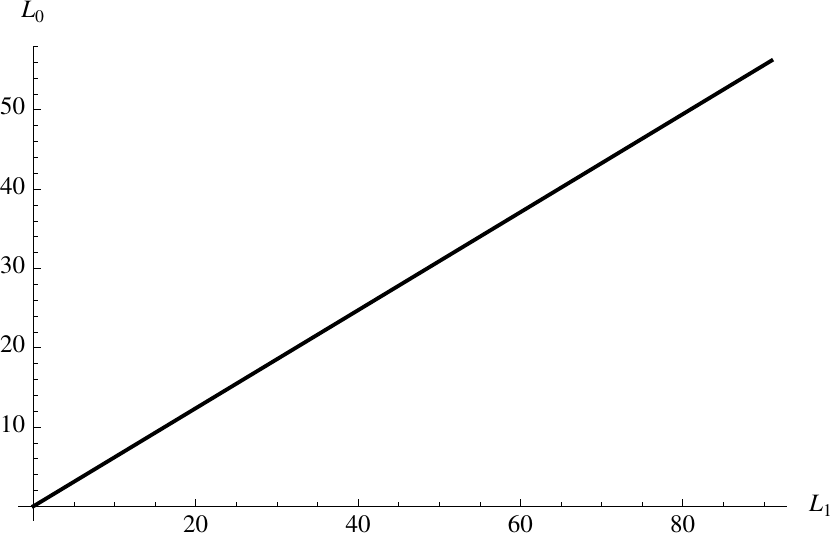}
\end{tabular}
\caption{Holographic mutual information in $AdS_4$ with $L_2=L_1$. On the left we show $M_2(L_1,L_1;L_0)$ for $L_0=0.87$ (red), $L_0=1.91$ (blue) and $L_0=3.93$ (black).
On the right, in the parameter space $(L_1,L_0)$, we plot the position of the transition point at which the mutual information starts to be non zero.
\label{2intM0free}}
\end{figure}

\subsection{$AdS_{d+2}$}
\label{AdS 2intervals}

For $AdS_{d+2}$ the analysis is simple because we explicitly know that (see (\ref{Atilde free}) and (\ref{alphad def}))
\begin{equation}
\label{Ad free 2int}
\tilde{A}_d(L) \,=\,-\,\frac{\alpha_d}{L^{d-1}}
\hspace{2cm}
\alpha_d =  \frac{1}{d-1}\left(\frac{2\sqrt{\pi}\, \Gamma\left(\frac{d+1}{2d}\right)}{\Gamma\left(\frac{1}{2d}\right)}\right)^d
\end{equation}
which holds for $d \geqslant 2$.
Keeping the distance $L_0$ between the two equal strips fixed, for small $L_1$ the pair of surfaces enclosing the disconnected volumes is minimal and the mutual information (\ref{MI def}) is zero. Increasing $L_1$, at a certain point the pair of surfaces enclosing the connected volume becomes minimal and the mutual information (\ref{MI def}) is therefore positive. For large $L_1$ the mutual information goes asymptotically to a constant, as shown for $d=2$ in the figure \ref{2intM0free} (plot on the left).
In order to find the asymptotic value of the mutual information, we observe from (\ref{Ad free 2int}) that $\tilde{A}_d(L)\rightarrow 0$ when $L \rightarrow \infty$. This implies that 
\begin{equation}
\lim_{L_1 \rightarrow \infty} M_d(L_1,L_1;L_0) \,=\, -\tilde{A}_d(L_0)\;.
\end{equation}
which provides the asymptotic value of the mutual information as a function of the distance between the strips.
\begin{figure}[h]
\begin{center}
\includegraphics[width=10cm]{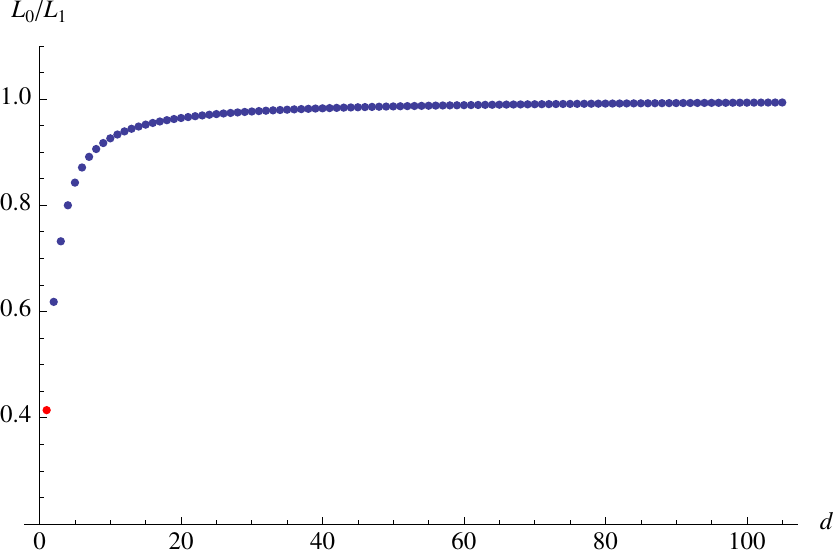}
\end{center}
\caption{Two equal and parallel strips. 
Angular coefficient of the line characterizing the transition of the holographic mutual information in $AdS_{d+2}$ in terms of $d$. The red point corresponds to $AdS_3$, which is not described by the equation (\ref{AdS transition eq}).
In this case the transition occurs at the value $x=1/2$ of the conformal ratio \cite{Headrick:2010zt}.
\label{free case transition}}
\end{figure}

\noindent As for the transition point at which the mutual information starts to be positive, its defining relation (\ref{transition eq}) specified for $AdS_{d+2}$ and $L_2=L_1$ reads
\begin{equation}
\label{AdS transition eq}
(L_0/L_1+2)^{d-1}\,=\,\frac{(L_0/L_1)^{d-1}}{2(L_0/L_1)^{d-1}-1}
\hspace{2cm} d\,\geqslant 2\,
\end{equation}
where (\ref{Ad free 2int}) has been employed. For any fixed $d\geqslant 2$, we can easily observe through a graphical analysis  that the equation (\ref{AdS transition eq}) has only one positive root for $L_0/L_1$. This root provides the angular coefficient of the straight line in the plane $(L_1,L_0)$.
In the figure \ref{2intM0free} (plot on the right) the case of $AdS_4$ is considered.\\
In the figure \ref{free case transition} we show the angular coefficient of the straight line, namely the solution of (\ref{AdS transition eq}),  as function of $d$.
We remark that the equation (\ref{AdS transition eq}) holds for $d \geqslant 2$. 
The case of $AdS_3$ (i.e. $d=1$) has been studied in \cite{Headrick:2010zt}, finding that the transition occurs when the conformal ratio $x\equiv z_{12} z_{34}/(z_{13} z_{24})=L_1^2/(L_1+L_0)^2=1/2$, which corresponds to the red point in the figure \ref{free case transition}.

\subsection{Charged black holes}
\label{section 2 int charged bh}

In this section we consider the holographic mutual information for a charged black hole. \\
By employing the results of the section \ref{section near horizon}, we have that
\begin{equation}
\label{Ad tilde BH}
\tilde{A}_d(L) \,=\,\frac{L}{z_0^d}+ c_d+o(1)
\hspace{2cm}
\textrm{for large $L$}
\end{equation}
where $c_d$ is the $O(1)$ term in (\ref{expansion nh shwarz}), (\ref{expansion nh rn T0}) and (\ref{expansion nh rn T}). Since we are not able to determine $c_d$ analytically, we fix it by fitting the numerical values of $\tilde{A}_d(L)$ at large $L$ with a line.
\begin{figure}[h]
\begin{tabular}{ccc}
\includegraphics[width=7.5cm]{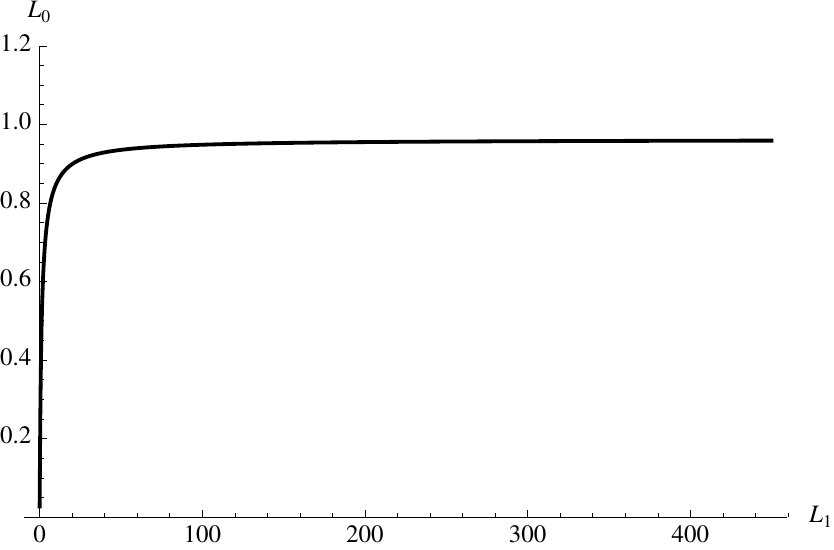}
& & 
\includegraphics[width=7.5cm]{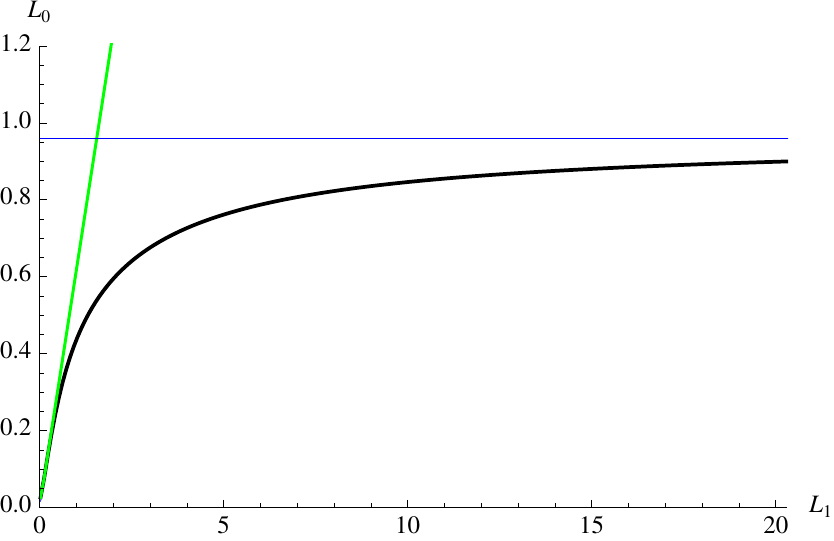}
\end{tabular}
\caption{Extremal charged black hole in $AdS_4$ with $z_0=1$. 
On the left the position of the transition point of the holographic mutual information in the parameter space $(L_1,L_0)$. On the right, a zoom of the same plot: the asymptotic line is provided by the equation (\ref{trans eq asympt L0}) and the green line corresponds to the transition point of $AdS_4$ (figure \ref{2intM0free}, plot on the right).
\label{2intBHd2trans}}
\end{figure}

\begin{figure}[h]
\begin{center}
\includegraphics[width=10cm]{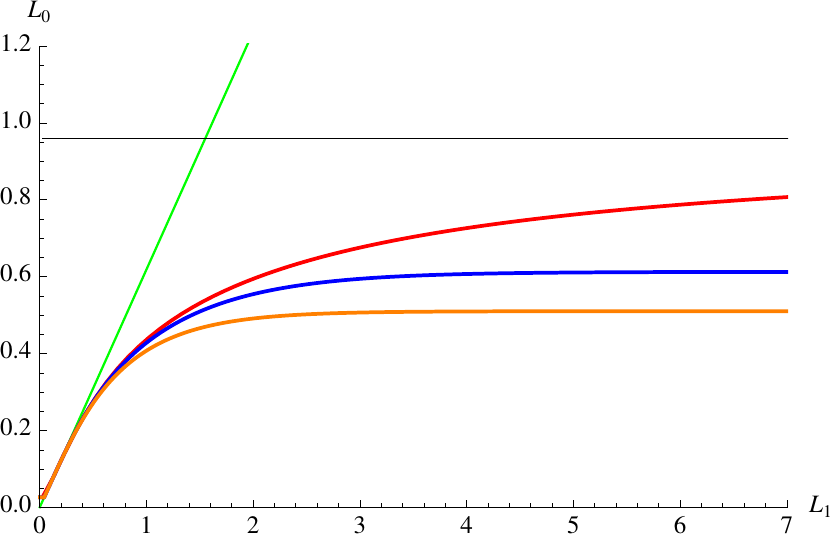}
\end{center}
\caption{Charged black hole in $AdS_4$ at fixed charge.  
Position of the transition point for $M_2(L_1,L_1;L_0)$ in the plane $(L_1,L_0)$: the red curve corresponds to the extremal case (see the figure \ref{2intBHd2trans}, plot on the right) while the blue and the orange ones correspond to two non extremal cases ($T=0.1$ and $T=0.18$ respectively). The black horizontal line gives the asymptotic value of the extremal case while the green line corresponds to the transition point of $AdS_4$.
\label{chargedBH trans comparison}}
\end{figure}

\noindent For two equal strips of width $L_1$ at fixed distance $L_0$, the behavior of the mutual information is qualitatively the same obtained for $AdS_{d+2}$ and shown in the figure \ref{2intM0free} (plot on the left). The asymptotic value of $M_d(L_1,L_1;L_0)$ at fixed $L_0$ can be found by employing (\ref{Ad tilde BH}). For $L_1 \rightarrow \infty$ we have
\begin{equation}
M_d(L_1,L_1;L_0) 
\,=\,  2\tilde{A}_d(L_1)-\tilde{A}_d(L_0)-\tilde{A}_d(2L_1+L_0)
\,\longrightarrow\,
c_d - \tilde{A}_d(L_0)-\frac{L_0}{z_0^d}\;.
\end{equation}
As for the position of the transition point of $M_d(L_1,L_1;L_0) $ in the plane $(L_1,L_0)$, the curve is instead qualitatively different from the corresponding one obtained for $AdS_{d+2}$,
Indeed, while we get a straight line for $AdS_{d+2}$ (plot on the right in the figure \ref{2intM0free}), for the charged black hole we find a curve with an asymptotic constant value (plot on the left in the figure \ref{2intBHd2trans}).
In particular, the straight line of $AdS_{d+2}$ is tangent to the curve corresponding to the charged black hole which is asymptotically $AdS_{d+2}$, as shown by the plot on the right in the figure \ref{2intBHd2trans}. Indeed, for small values of $L_1$ the pairs of surfaces to compare are close to the boundary and consequently the transition between them is determined by the asymptotic geometry.\\
Let us consider further the characteristic asymptotic value $\tilde{L}_0$ of the curve of the transition points of the mutual information for a charged black hole in the plane $(L_1,L_0)$ as $L_1$ becomes large.
The equation defining $\tilde{L}_0$ can be found by taking the limit $L_1 \rightarrow \infty$ and $L_0 \rightarrow \tilde{L}_0$ of the equation (\ref{transition eq}) and employing (\ref{Ad tilde BH}). The result is
\begin{equation}
\label{trans eq asympt L0}
 \tilde{A}_d(\tilde{L}_0)+\frac{\tilde{L}_0}{z_0^d}-c_d\,=\,0
\end{equation}
which can be solved numerically. 
This asymptotic value of the distance between the two strips could be interpreted as a signal of the occurrence of a finite correlation length in the boundary theory.\\
The qualitative features just described for the extremal charged black are found for the non extremal case as well. The mutual information $M_d(L_1,L_1;L_0)$ behaves like in the plot on the left of the figure \ref{2intM0free} and the curve of the transition points is qualitatively like the one shown in the figure \ref{2intBHd2trans}, with the asymptotic value given by the solution of the equation (\ref{trans eq asympt L0}) with the proper emblacking function depending on the temperature.
In the figure \ref{chargedBH trans comparison} we show the curves of transition points of $M_2(L_1,L_1;L_0)$ for two different temperatures besides the extremal case at fixed charge. 
The curve corresponding to a certain temperature always stays below the curve corresponding to a lower temperature, meaning that the asymptotic value determined by (\ref{trans eq asympt L0}) decreases with the temperature for a fixed charge of the black hole.
We recall that imposing $Q$ fixed implies that we cannot change the temperature keeping fixed the position of the horizon $z_0$ because these quantities are related through (\ref{eq Tz0 fixed Q}).

\section{Conclusions}

In this paper we have considered two aspects of the holographic entanglement entropy in black hole backgrounds: the behavior of the finite term as the width of the strip in the boundary becomes large and the transition of the mutual information for two equal strips in the parameter space given by the width of the strips and their distance.\\
For one strip in the limit of large volume, which means that the turning point of the minimal surface approaches the horizon, we confirm and extend to new cases the known result that the finite term scales like the width, and therefore like the volume, of the strip. The distinguished feature of the different black holes is the degree of the divergence of the finite part in terms of the distance between the turning point and the horizon, which is determined by the near horizon geometry. In the case of a Lifshitz background with a dynamical exponent entering in the spatial part of the metric, such scaling could be influenced by this exponent.
For a Lifshitz black hole in four dimensions we computed the analytic expression of the holographic entanglement entropy to all orders in the UV cutoff.\\
For two equal and parallel strips in the boundary, we have found that the transition of the mutual information for a charged black hole naturally provides a finite limiting distance between the strips as their width becomes large. This asymptotic value could be interpreted as a signal of a finite correlation length in the boundary theory. The transition in the mutual information is characteristic of the holographic prescription; therefore it is a large $c$ effect. We believe that it is important to further study this transition in order to understand how it smooths out for finite $c$. 
This is part of the general aim of reproducing through holography the results obtained for the mutual information in the finite $c$ models.

\subsection*{Acknowledgments}

It is a pleasure to thank Hong Liu for collaboration in the initial stage of this project and for many helpful discussions and advices during its development.
I am also grateful to Matthew Headrick, Mark Hertzberg, Veronika Hubeny, John McGreevy, Mukund Rangamani, Tadashi Takayanagi, Frank Wilczek and in particular to Antonello Scardicchio for useful discussions.
I thank the Physics department of the University of Pisa and the Galileo Galilei Institute for their kind hospitality during the last part of this project.\\
This work is mainly supported by Istituto Nazionale di Fisica Nucleare (INFN) through a Bruno Rossi fellowship and also by funds of the U.S. Department of Energy (DoE) under the cooperative research agreement DE-FG02-05ER41360.

\appendix

\section{$AdS_{d+2}$}
\label{app AdS}

For the sake of completeness, in this appendix we briefly review the results for the holographic entanglement entropy in $AdS_{d+2}$ for the strip \cite{Ryu:2006bv, Ryu:2006ef}. The expressions in the section \ref{section strip BH f(z)} can be applied with $f(z)=1$ identically.\\
The inverse function of the profile $z(x)$ representing the minimal surface is given by
\begin{eqnarray}
x(z)&=&
\int_z^{z_{\textrm{max}}}
\frac{w^d}{\sqrt{z_{\textrm{max}}^{2d}-w^{2d}}}\,dw
\;=\;
\frac{w^{d+1}}{(d+1) z_{\textrm{max}}^d}\;
_2 F_1\left(\frac{d+1}{2d}\, ,\frac{1}{2}\, ; \frac{3d+1}{2d}\, ;\frac{w^{2d}}{z_{\textrm{max}}^{2d}}\right)
\Bigg|^{z_{\textrm{max}}}_z
\hspace{1cm}
\\
\label{free profile}
\rule{0pt}{.8cm}&=&
\frac{\sqrt{\pi}\,\Gamma\left(\frac{d+1}{2d}\right)}{\Gamma\left(\frac{1}{2d}\right)}\,z_{\textrm{max}}-
\frac{z^{d+1}}{(d+1) z_{\textrm{max}}^d}\;
_2 F_1\left(\frac{d+1}{2d}\, ,\frac{1}{2}\, ; \frac{3d+1}{2d}\, ;\frac{z^{2d}}{z_{\textrm{max}}^{2d}}\right)\;.
\end{eqnarray}
Since $x(0)=L/2$, from (\ref{free profile}) we see that
\begin{equation}
\label{L free}
L\,=\,
\frac{2\sqrt{\pi}\,\Gamma\left(\frac{d+1}{2d}\right)}{\Gamma\left(\frac{1}{2d}\right)}\,z_{\textrm{max}}\;.
\end{equation}
As for the regularized area of this minimal surface, it is given by (\ref{strip area}) where now the integral to perform is 
\begin{eqnarray}
A_d(z_{\textrm{max}},a)
&=&
2\int_a^{z_{\textrm{max}}}
\frac{z_{\textrm{max}}^{d}}{w^d\sqrt{z_{\textrm{max}}^{2d}-w^{2d}}}\,dw
\\
\label{area free step3}
\rule{0pt}{.8cm}&=&
 \frac{2}{(d-1)\,a^{d-1}}- \frac{2}{(d-1)\,z_{\textrm{max}}^{d-1}}
 + \int_a^{z_{\textrm{max}}} 
\frac{2}{w^d}\left(\frac{z_{\textrm{max}}^d}{\sqrt{z_{\textrm{max}}^{2d}-w^{2d}}}-1\right)dw
\hspace{.7cm}
\end{eqnarray}
where the divergence for small $a$ has been isolated as in (\ref{split 0}) or (\ref{split 1}) (in absence of the black hole they provide the same result). The integral in (\ref{area free step3}) reads
\begin{eqnarray}
\hspace{0cm} 
\int_a^{z_{\textrm{max}}} 
\frac{2}{w^d}\left(\frac{z_{\textrm{max}}^d}{\sqrt{z_{\textrm{max}}^{2d}-w^{2d}}}-1\right)dw
\;=
& & \\
\rule{0pt}{.8cm}& &\hspace{-6.5cm} =\;
\left[\,
\frac{2}{(d-1)\,w^{d-1}}\left(1-\sqrt{1-\frac{w^{2d}}{z_{\textrm{max}}^{2d}}}\,\right)
-\frac{2w^{d+1}}{(d^2-1)z_{\textrm{max}}^{2d}}\;
_2 F_1\left(\frac{d+1}{2d}\, ,\frac{1}{2}\, ; \frac{3d+1}{2d}\, ;\frac{w^{2d}}{z_{\textrm{max}}^{2d}}\right)
\right]\Bigg|_{a}^{z_{\textrm{max}}}
\nonumber \\
& &\hspace{-6.5cm} =\;
\frac{2}{(d-1)z_{\textrm{max}}^{d-1}}-\frac{\sqrt{\pi}\;\Gamma\big(\frac{d+1}{2d}\big)}{(d-1)\,\Gamma\big(\frac{1}{2d}\big)\,z_{\textrm{max}}^{d-1}}
+ O(a^{d+1})
\hspace{1.6cm} d\,\geqslant\,2\;.
\end{eqnarray}
Thus the UV divergence of $\textrm{Area}(\gamma_A)$ has been isolated and
the final result is \cite{Ryu:2006ef}
\begin{eqnarray}
\label{A expansion empty}
A_d(z_{\textrm{max}},a) &=&
 \frac{2}{(d-1)\,a^{d-1}}
-\frac{2\sqrt{\pi}\;\Gamma\big(\frac{d+1}{2d}\big)}{(d-1)\,\Gamma\big(\frac{1}{2d}\big)\,z_{\textrm{max}}^{d-1}} + O(a^{d+1})\\
\label{Atilde free}
&=&
 \frac{2}{(d-1)\,a^{d-1}}-\frac{\alpha_d}{L^{d-1}}
 + O(a^{d+1})
\end{eqnarray}
where
\begin{equation}
\label{alphad def}
 \alpha_d \,\equiv\,
 \frac{1}{d-1}\left(\frac{2\sqrt{\pi}\, \Gamma\left(\frac{d+1}{2d}\right)}{\Gamma\left(\frac{1}{2d}\right)}\right)^d\;.
\end{equation}
This expression has been employed in the section \ref{AdS 2intervals} to study the asymptotic value of the mutual information.\\
Now we find it useful to derive (\ref{A expansion empty}) also in the following way, which could be employed in a generalized version for the black holes. 
First one writes the integral in (\ref{area free step3}) as a series
\begin{eqnarray}
\label{empty area 1}
\int_a^{z_{\textrm{max}}} 
\frac{2}{w^d}\left(\frac{z_{\textrm{max}}^d}{\sqrt{z_{\textrm{max}}^{2d}-w^{2d}}}-1\right)dw
&=&
\sum_{n = 1}^{\infty}
\frac{2b_n}{z_{\textrm{max}}^{2d n}}
\int_a^{z_{\textrm{max}}} w^{d(2n-1)}dw\\
\label{fin term exp}
& = &
\frac{2}{z_{\textrm{max}}^{d-1}}
\sum_{n = 1}^{\infty} \frac{b_n}{(2n-1)d+1}
+O(a^{d+1})
\end{eqnarray}
where in (\ref{empty area 1}) the coefficients $b_n$ can be found by employing the following identity with $\alpha=1/2$
\begin{equation}
\label{b_n}
\frac{1}{(1-x)^{\alpha}}=\,
_2 F_1(\alpha,\beta;\beta;x)\,=\,
\sum_{n=0}^{\infty} \frac{(\alpha)_n}{n!}\,x^n
\hspace{1cm}\Longrightarrow \hspace{1cm}
b_n\,=\,\frac{(1/2)_n}{n!}
\end{equation}
being $(c)_n\equiv c(c+1)\dots (c+n-1)$ the Pochhammer symbol (we recall that $(c)_0\equiv 1$). \\
Then, from (\ref{area free step3}) and (\ref{fin term exp}) we get that for the finite term in the expansion for small $a$
\begin{equation}
\frac{2}{z_{\textrm{max}}^{d-1}}
\left(-\,\frac{1}{d-1}+
\sum_{n = 1}^{\infty} \frac{b_n}{(2n-1)d+1}
\right)
\,=\,
-\,\frac{2\sqrt{\pi}\;\Gamma\big(\frac{d+1}{2d}\big)}{(d-1)\,\Gamma\big(\frac{1}{2d}\big)\,z_{\textrm{max}}^{d-1}} 
\end{equation}
which agrees with the finite term in (\ref{A expansion empty}).

\section{Charged black holes in $AdS_{d+2}$}
\label{app chargedBH}

In this appendix we review some features of the charged black holes which are asymptotically $AdS_{d+2}$.
The metric reads
\begin{equation}
\label{metric charged BH}
ds^2\,=\,\frac{r^2}{R^2}\big( - f dt^2+d\vec{x}^{2}\big)+\frac{R^2}{r^2}\,\frac{dr^2}{f}
\hspace{2cm}
f\,=\,1+\frac{Q^2}{r^{2d}}-\frac{M}{r^{d+1}}
\end{equation}
for $d \geqslant 2$, where $d\vec{x}^2$ is the metric of $\mathbb{R}^{d}$, $M$ is the mass and $Q$ is the charge of the black hole. 
The boundary corresponds to large $r$, where the metric becomes the one of $AdS_{d+2}$ with radius $R$.
The Schwarzschild black hole in $AdS_{d+2}$ is obtained by setting $Q=0$.\\
By introducing the variable $z \equiv R^2/r$, the metric (\ref{metric charged BH}) becomes (\ref{charged BH metric z}) and the boundary corresponds to $z=0$. This parameterization is largely used in this paper.
Another useful parameterization of the radial coordinate is 
\begin{equation}
e^{\tilde{r}/R} \,=\,\frac{r}{R}\;.
\end{equation}
Notice that a scaling of $r$ corresponds to a shift of $\tilde{r}$. With the parameterization given by $\tilde{r}$, 
the metric (\ref{metric charged BH}) reads
\begin{equation}
ds^2\,=\,e^{\tilde{r}/R} 
\big[ - f(\tilde{r}) dt^2+d\vec{x}^{2}\,\big]+\frac{d\tilde{r}^2}{f(\tilde{r})}
\hspace{.9cm}
f(\tilde{r})\,=\,1+\frac{Q^2}{R^{2d}}\,e^{-2d \,\tilde{r}/R}-\frac{M}{R^{d+1}}
\,e^{-(d+1)\tilde{r}/R}\;.
\end{equation}
It is convenient to parameterize $Q$ by introducing $r_\ast$ as follows
\begin{equation}
Q^2 \equiv\, \frac{d+1}{d-1}\,r_\ast^{2d}\,=\,\frac{d+1}{d-1}\left(\frac{R^2}{z_\ast}\right)^{2d}\;.
\end{equation}
From this expression it is evident that $Q$ has the dimension of $[L]^d$.
The limit $z_\ast \rightarrow \infty$ corresponds to the Schwarzschild black hole.
The chemical potential reads
\begin{equation}
\mu\, \equiv\, 
\sqrt{\frac{d}{2(d-1)}}\,\frac{g_{F}Q}{R^2 r_0^{d-1}}
\,=\,
\frac{\sqrt{d(d+1)}}{\sqrt{2}(d-1)}\,\frac{g_F r_0}{R^2}\left(\frac{r_\ast}{r_0}\right)^{d}
\,=\,
\frac{\sqrt{d(d+1)}}{\sqrt{2}(d-1)}\,\frac{g_F}{z_0}\left(\frac{z_0}{z_\ast}\right)^{d}
\end{equation}
where $g_F$ is the effective dimensionless gauge coupling.
When $z_\ast \rightarrow \infty$ for fixed $z_0$ the chemical potential $\mu$ vanishes.
The temperature is 
\begin{equation}
\label{temp}
T\,=\,\frac{(d+1)r_0}{4\pi R^2}\left(1-\frac{r_\ast^{2d}}{r_0^{2d}}\right)
\,=\,\frac{d+1}{4\pi z_0}\left(1-\frac{z_0^{2d}}{z_\ast^{2d}}\right)
\,=\,\frac{d+1}{4\pi z_0}\left(1-\frac{d-1}{d+1}\,\frac{Q^2 z_0^{2d}}{R^{4d}}\right)
\,\geqslant\,0\;.
\end{equation}
Since $r_0 \geqslant r_\ast$ in order to impose $T>0$, we have $z_0 \leqslant z_\ast$.
Notice that if we want to keep $Q$ fixed, changing $T$ implies a change of $z_0$. Indeed, the values of $Q$ and $T$ fix the position $z_0$ of the horizon through (\ref{temp}), which can be written also as follows
\begin{equation}
\label{eq Tz0 fixed Q}
\frac{(d-1)Q^2}{(d+1)R^{4d}}\, z_0^{2d}+\frac{4\pi T}{d+1}\, z_0-1 \,=\,0\;.
\end{equation}
Setting $R=1$, if we decide to choose $z_0=1$ at $T=0$ then $Q^2=(d+1)/(d-1)$.
Keeping this value for $Q^2$ fixed, moving to $T>0$ modifies $z_0$ according to (\ref{temp}) which becomes
\begin{equation}
z_0^{2d}+\frac{4\pi T}{d+1}\, z_0-1 \,=\,0\;.
\end{equation}
From the relation (\ref{temp}) it seems that there is a maximum temperature corresponding to $z_\ast \rightarrow \infty$.
Instead the relevant parameter is the ratio
\begin{equation}
\label{T/mu}
\frac{T}{\mu}\,=\,\frac{\sqrt{2(d+1)}(d-1)z_\ast^d}{4\pi g_F\sqrt{d} \,z_0^d}
\left(1-\frac{z_0^{2d}}{z_\ast^{2d}}\right)
\,\equiv\,
\tilde{\alpha}_d
\left[ 
\left(\frac{z_\ast}{z_0}\right)^{d}-\left(\frac{z_\ast}{z_0}\right)^{-d}\,
\right]
\hspace{.6cm}
\tilde{\alpha}_d \equiv \frac{\sqrt{2(d+1)}(d-1)}{4\pi g_F\sqrt{d}}
\end{equation}
which spans all the positive real numbers when $z_\ast \in [z_0,\infty)$ in a strictly monotonical way, going to infinity when $z_\ast \rightarrow \infty$. From (\ref{T/mu}) we can see that (the other root is negative)
\begin{equation}
\label{zast/z0}
\left(\frac{z_\ast}{z_0}\right)^{d}=\,
\frac{T}{2\tilde{\alpha}_d\mu}+\sqrt{\left(\frac{T}{2\tilde{\alpha}_d\mu}\right)^2+1}\,\geqslant\,1
\end{equation}
which  becomes $1$ when $T=0$ for any $d \geqslant 2$.
The parameter $M$, which can be expressed in terms of $Q^2$ and the position $r_0$ of the horizon, reads
\begin{equation}
M\,=\,r_0^{d+1}+\frac{Q^2}{r_0^{d-1}}\,=\,r_0^{d+1}+\frac{d+1}{d-1}\,\frac{r_\ast^{2d}}{r_0^{d-1}}
\,=\,\left(\frac{R^2}{z_0}\right)^{d+1}\left[\,1+\frac{d+1}{d-1}\left(\frac{z_0}{z_\ast}\right)^{2d}\,\right]\;.
\end{equation}
Thus, the emblacking function can be written as follows
\begin{eqnarray}
\label{f(z) general}
f(z) &=&
1+\frac{d+1}{d-1}\left(\frac{z}{z_\ast}\right)^{2d}
-\left[\,1+\frac{d+1}{d-1}\left(\frac{z_0}{z_\ast}\right)^{2d}\,\right]\left(\frac{z}{z_0}\right)^{d+1}\\
&=&
\label{f charged bh}
1+\frac{d+1}{d-1}\left(1-\frac{4\pi z_0}{d+1}\,T\right)\left(\frac{z}{z_0}\right)^{2d}
-\frac{2d}{d-1}\left(1-\frac{4\pi z_0}{2d}\,T\right)
\left(\frac{z}{z_0}\right)^{d+1}\;.
\end{eqnarray}
Notice that from (\ref{f(z) general}) and (\ref{zast/z0}) we can write $f(z)$ in terms of the ratio $T/\mu$.
A very important role in our discussions is recovered by the near horizon geometry, namely the one obtained when $z \rightarrow z_0$. 
Close to the horizon, the emblacking function can be expanded as
\begin{eqnarray}
f(z)&=&
\frac{(d+1)(z_\ast^{2d}-z_0^{2d})}{z_0}(z-z_0)
+\frac{d(d+1)(3z_0^{2d}-z_\ast^{2d})}{2z_0^2}(z-z_0)^2
+O\big((z-z_0)^3\big)\hspace{.5cm}\\
\label{f(z) nh T}
\rule{0pt}{.7cm}&=&
4\pi T (z_0-z)+\frac{d(d+1-6\pi z_0 T)}{z_0^2}(z-z_0)^2+O\big((z-z_0)^3\big)\;.
\end{eqnarray}
In the extremal case ($T=0 \Leftrightarrow z_\ast=z_0$) the emblacking function 
$f(z)=O \big((z-z_0)^2\big)$, while in the non extremal case ($T>0$ and $z_\ast > z_0$) we have $f(z)=O(z-z_0)$ when $z \rightarrow z_0$.
We remark that also in the case of the Schwarzschild black hole, which corresponds to $z_\ast \rightarrow \infty$, we have $f(z)=O(z-z_0)$ as $z \rightarrow z_0$.

\section{Disk geometry}
\label{app circle}

In this appendix we briefly discuss the case in which the region $A$ in the spatial section of the boundary theory is given by a disk, while
in the bulk a black hole occurs whose metric on the constant time slice is given by (\ref{ds20 ansatz}).\\
Taking as $A$ the circle given by  $\rho=\widetilde{R}$ (it is more convenient to adopt the polar coordinates) and assuming that $z=z(\rho)$, we get
\begin{equation}
\label{disk area functional}
\textrm{Area}(\gamma_A)\,=\,
V_{d-1} \int_0^{\widetilde{R}}
d\rho \,\rho^{d-1} \left(\frac{R}{z}\right)^{d}\sqrt{1+\frac{(z')^2}{f(z)}}
\end{equation}
where $z'=dz/d\rho$ and $V_{d-1}$ is the volume of the $d-1$ unit sphere.
Now the Lagrangian density $\mathcal{L}_{\textrm{disk}}[z(\rho)]$ is the integrand of (\ref{disk area functional})
and it explicitly depends on the coordinate $\rho$. This means that there is not a conserved first integral.\\ 
In order to minimize the functional (\ref{disk area functional}) we need to solve the second order equation given by the equation of motion, which is
\begin{equation}
\label{eq diff disk}
\rho\,\frac{z'' z}{f(z)}
+(d-1)\,\frac{z'\,z}{f(z)} \left[1+\frac{(z')^2}{f(z)}\right]
+d\,\rho\left[1+\frac{(z')^2}{f(z)}\right]
- \rho\,\frac{(z')^2 z}{2 f(z)^2}\,f'(z)
\,=\,0
\end{equation}
where $f'(z)=df(z)/dz$.
Thus, this case is more complicated than the strip, largely considered throughout the paper, because now we have to solve a second order equation to find the profile to use in the integral giving the area.\\
For $AdS_{d+2}$ the equation to solve is (\ref{eq diff disk}) with $f(z)=1$ identically (see the footnote 20 of \cite{Ryu:2006ef}) and its solution reads
\begin{equation}
z_0(\rho)\,=\,\sqrt{ \widetilde{R}^2-\rho^2}
\end{equation}
which is the semispherical surface whose $A$ is the maximal circle. For a black hole background, which has a non trivial emblacking function $f(z)$, the equation (\ref{eq diff disk}) for the profile of the minimal surface can be solved numerically.

\section{An alternative splitting of the finite term}
\label{app more on nh}

In this appendix we provide some insights about the expansion for $z_{\textrm{max}} \rightarrow z_0$ of the finite term of the holographic entanglement entropy and about the role of the near horizon geometry by considering the splitting (\ref{split 1}).\\
Let us assume to know the first integral in (\ref{I def}) analytically. Then, the $O(1)$ term of $A_d(z_{\textrm{max}},a)$ in the expansion for $a \rightarrow 0$ is obtained by $I_d(0, z_{\textrm{max}})$ plus a contribution from the first integral.
In general we are unable to compute $I_d(0, z_{\textrm{max}})$.
Anyway, we are interested into its expansion as $z_{\textrm{max}} \rightarrow z_0$.
The emblacking function $f(w)$ depends on the ratio $w/z_0$. By introducing $y \equiv w/z_{\textrm{max}}  \in [0,1]$ as integration variable, the function  $f(z_{\textrm{max}}\, y)$ depends on the ratio $z_{\textrm{max}} /z_0 < 1$, therefore we can consider the expansion of the function $1/\sqrt{f(z_{\textrm{max}}\, y)}$ as $z_{\textrm{max}} /z_0 \rightarrow 1^-$, obtaining
\begin{eqnarray}
\label{Id y var}
I_d(0, z_{\textrm{max}})
&=&
\frac{2}{z_{\textrm{max}}^{d-1}}
\int_0^{1} 
\frac{1}{y^d \sqrt{f(z_{\textrm{max}} \,y)}}
\left(\frac{1}{\sqrt{1-y^{2d}}}-1\right) dy
\\
\label{Id maybe expanded}
\rule{0pt}{.8cm}
& \equiv &
\frac{2}{z_{\textrm{max}}^{d-1}}
\int_0^{1} \sum_{n=0}^{\infty}
\frac{h_n(y)}{y^d}
\left(\frac{1}{\sqrt{1-y^{2d}}}-1\right) 
\left(1- \frac{z_{\textrm{max}}}{z_0}\right)^n
dy\;.
\end{eqnarray}
Unfortunately, the integral and the series cannot be inverted because the integrals
occurring for any fixed $n$ are divergent at the upper extremum $y=1$ as we will see below in a special case. 
By  introducing an intermediate scale $a < z_\lambda < z_{\textrm{max}}$, we can write
\begin{equation}
\label{splitting}
I_d(0, z_{\textrm{max}})\,=\,
I_d(0, z_\lambda)+I_d(z_\lambda, z_{\textrm{max}})\;.
\end{equation}
Now, in $I(0, z_\lambda)$ we can invert the series and the integral because the upper limit is $z_\lambda/z_{\textrm{max}} <1$ and the integrals converge. We get
\begin{equation}
\label{I exp term1}
I_d(0, z_\lambda)\,=\,
\frac{2}{z_{\textrm{max}}^{d-1}}
 \sum_{n=0}^{\infty}
 \left[\,
\int_0^{\frac{z_\lambda}{z_{\textrm{max}}}} 
\frac{h_n(y)}{y^d}
\left(\frac{1}{\sqrt{1-y^{2d}}}-1\right) dy \right]
\left(1- \frac{z_{\textrm{max}}}{z_0}\right)^n
\end{equation}
which is a well defined expansion whose coefficients depend on the ratio $z_\lambda/z_{\textrm{max}}$.\\
The second integral $I(z_\lambda, z_{\textrm{max}})$ is still divergent when $z_{\textrm{max}} /z_0 \rightarrow 1^-$ and we cannot invert the series with the integration as done in (\ref{I exp term1}); therefore it must be computed analytically. Since this is usually too difficult, we can approximate it by employing the near horizon behavior of the emblacking function. The closer is $z_\lambda$ to $z_{\textrm{max}}$, the better is this approximation.\\
In order to apply these considerations to a concrete example, let us consider the extremal charged black hole in $AdS_4$. The first integral in (\ref{I def}) in this case can be computed, obtaining
\begin{eqnarray}
\int_a^{z_{\textrm{max}}} 
\frac{2}{w^2 \sqrt{f(w)}}\,dw
&=&\\
& & \hspace{-2.6cm}
=\;
2\left[\,\frac{1}{\sqrt{6}z_0}
\log\left(\frac{4w+2z_0+\sqrt{6(3w^2+2z_0w+z_0^2)}}{z_0-w}\,\right)
-\frac{\sqrt{3w^2+2z_0w+z_0^2}}{z_0 w}
\,\right]\Bigg|_a^{z_{\textrm{max}}} 
\nonumber\\
\label{I2 integ1}
\rule{0pt}{.7cm}
& & \hspace{-2.6cm}
=\;\frac{2}{a}+\left[-\frac{2}{\sqrt{6} z_0} \log(z_0-z_{\textrm{max}}) +O(1)\right] + O(a^2)
\end{eqnarray}
where the square brackets in (\ref{I2 integ1}) enclose the finite term in the power series in $a$, which has been further expanded for $z_{\textrm{max}} \rightarrow z_0$. \\
Now, by expanding the integral in (\ref{I def}) as explained in the section \ref{section near horizon} we find
\begin{equation}
\label{I2 divergence charged BH}
I_2(z_{\textrm{max}},0)\,=\,
\frac{\pi}{\sqrt{6 z_0}\,\sqrt{z_0-z_{\textrm{max}}}}
+\frac{2}{\sqrt{6} z_0} \log(z_0-z_{\textrm{max}})
+ O(1)
\end{equation}
where, again, we do not control the finite term. Notice that the logarithmic divergence in (\ref{I2 divergence charged BH}) cancels the one in (\ref{I2 integ1}) and the remaining divergence is the same one found in (\ref{expansion nh rn T0}) by using (\ref{split 0}). This is a consistency check of the two splittings (\ref{split 0}) and (\ref{split 1}) of the same integral. \\
As discussed above in this appendix, let us consider the integral $I_2(0,z_{\textrm{max}})$ in terms of the variable $y$ (see (\ref{Id y var})).
The emblacking function then reads
\begin{equation}
f(z_{\textrm{max}} \,y)\,=\,
1 -4 \left(\frac{z_{\textrm{max}}}{z_0}\,y\right)^{3}
+3 \left(\frac{z_{\textrm{max}}}{z_0}\,y\right)^{4}\;.
\end{equation}
By expanding $1/\sqrt{f(z_{\textrm{max}} \,y)}$ for $z_{\textrm{max}}/z_0 \rightarrow 1^-$, we find the functions $h_n(y)$ occurring in the series (\ref{I exp term1}). For the first terms, they are e.g.
\begin{equation}
h_0(y)= \frac{1}{\sqrt{1-4y^3+3y^4}}
\hspace{1.4cm}
h_1(y) = -\,\frac{6(1-y)y^3}{(1-4y^3+3y^4)^{3/2}}
\hspace{1cm}\dots
\end{equation}
and the corresponding integrals obtained by inverting the summation  and the integration in (\ref{Id maybe expanded}) are divergent in 1 because $1-4y^3+3y^4=O((1-y)^2)$ when $y \rightarrow 1$.\\
\begin{figure}[h]
\begin{center}
\includegraphics[width=8.5cm]{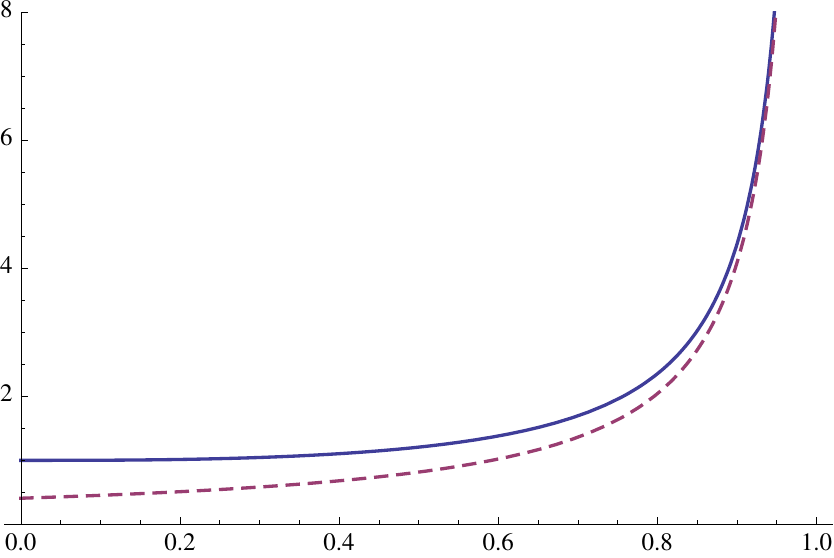}
\end{center}
\caption{Extremal charged black hole in $AdS_4$ and $z_0=1$. 
Plot of $1/\sqrt{f(z)}$ (solid line) and $1/\sqrt{f_\textrm{nh}(z)}$ (dashed line) as functions of $z \in (0,z_0)$.
\label{plot ffactors nh}}
\end{figure}
As discussed above, we introduce an intermediate scale $z_\lambda$ and split the integral as in (\ref{splitting}), obtaining for the first term a well defined power series (\ref{I exp term1}) in terms of integrals involving the functions $h_n$. We are not able to compute them analytically, but we are guaranteed that in $I_2(0, z_\lambda)$ is finite as $z_{\textrm{max}} \rightarrow z_0$. The divergence comes from the near horizon region.\\
As for the second integral in (\ref{splitting}) giving the divergent part for $z_{\textrm{max}} \rightarrow z_0$, we cannot compute it explicitly, but we can relate it to the corresponding integral involving the near horizon geometry. In particular, as shown in the figure \ref{plot ffactors nh}, the integral $I_2(z_\lambda, z_{\textrm{max}})$ is greater than the corresponding one computed with the near horizon geometry for any choice of $z_\lambda$; namely
\begin{equation}
\label{I2 > I2nh}
I_2(z_\lambda, z_{\textrm{max}})
\,>\,
I_{2,\textrm{nh}}(z_\lambda, z_{\textrm{max}})\,\equiv
\int_{z_\lambda}^{z_{\textrm{max}}} 
\frac{2}{w^2 \sqrt{f_{\textrm{nh}}(w)}}\left(\frac{z_{\textrm{max}}^2}{\sqrt{z_{\textrm{max}}^{4}-w^{4}}}-1\right) dw
\end{equation}
where the emblacking function close to the horizon reads (see (\ref{f(z) nh T}))
 \begin{equation}
f_{\textrm{nh}}(w)\,=\,6\,\frac{(w-z_0)^2}{z_0^2}\;.
\end{equation}
The integral in (\ref{I2 > I2nh}) is easier to deal with and the closer $z_\lambda$ is to $z_{\textrm{max}}$ the better is the approximation obtained by substituting $I_2(z_\lambda, z_{\textrm{max}})$ with $I_{2,\textrm{nh}}(z_\lambda, z_{\textrm{max}})$.


\end{document}